\documentclass[12pt,a4paper]{article}

\usepackage{jheppub}
\author[a]{Marc-Antoine Fiset,}
\author[a]{Matthias R. Gaberdiel}
\affiliation[a]{Institut f\"ur Theoretische Physik,
ETH Z\"urich,\\
Wolfgang-Pauli-Stra{\ss}e 27,
8093 Z\"urich, Switzerland}
\emailAdd{mfiset@phys.ethz.ch, gaberdiel@itp.phys.ethz.ch}

\usepackage{xcolor}
     \definecolor{green_maf}{RGB}{28, 166, 46}
     \definecolor{blue_mrg}{RGB}{12, 143, 145}
     \definecolor{detail}{RGB}{110,110,110}
     \definecolor{darkBlue}{RGB}{0,62,133}
     \definecolor{bcite}{RGB}{119, 150, 166}
\usepackage{setspace} 
\usepackage[normalem]{ulem} 
\usepackage{amsmath} 
\usepackage{nccmath} 
\usepackage{amssymb} 
\usepackage{amsthm} 
\usepackage{upgreek} 
\usepackage{simpler-wick} 
\usepackage{datetime} \settimeformat{ampmtime}
\usepackage[new]{old-arrows} 
\usepackage{dsfont} 
\usepackage{bm} 
\usepackage{pbox} 
\usepackage{tikz} 
\usepackage{colortbl} 
\usepackage{mathrsfs} 
\usepackage{mathtools} 
\usepackage[utf8]{inputenc} 
\usepackage{fancyhdr} 
\usepackage{changepage} 

\newif\ifcomments
     \commentstrue
\newif\ifdetails
     \detailstrue
\newif\ifmafdetails
     \mafdetailstrue

\DeclarePairedDelimiter\ket{\lvert}{\rangle} 
\DeclarePairedDelimiterX\braket[2]{\langle}{\rangle}{#1 \delimsize\vert #2} 

\def\be{\begin{equation}}
\def\ee{\end{equation}}

\begin{document}

\title{Deformed Shatashvili-Vafa algebra for superstrings on AdS$_3\times {\cal M}_7$}

\abstract{String backgrounds of the form $\mathbb{M}_3 \times {\cal M}_7$ where $\mathbb{M}_3$ denotes $3$-dimensional Minkowski space while ${\cal M}_7$ is a $7$-dimensional G$_2$-manifold, are characterised by the property that the world-sheet theory has
a Shatashvili-Vafa (SV) chiral algebra. We study the generalisation of this statement to backgrounds where the Minkowski factor $\mathbb{M}_3$ is replaced by ${\rm AdS}_3$. We argue that in this case the world-sheet theory is characterised by a certain ${\cal N}=1$ superconformal ${\cal W}$-algebra that has the same spin spectrum as the SV algebra and also contains a tricritical Ising model ${\cal N}=1$ subalgebra. We determine the allowed representations of this ${\cal W}$-algebra, and  analyse to which extent the special features of the SV algebra survive this generalisation.
}

\maketitle

\section{Introduction}

Some years ago Shatashvili and Vafa \cite{Shatashvili:1994zw} proposed that supersymmetric string backgrounds of the form 
\be\label{flat}
\mathbb{M}_3 \times {\cal M}_7  \ , 
\ee
where $\mathbb{M}_3$ denotes $3$-dimensional Minkowski space, while ${\cal M}_7$ is a $7$-dimensional G$_2$-manifold\footnote{We use `G$_2$-manifold' quite loosely to mean `manifold with a G$_2$-structure'. The $7$d geometry should support a nowhere-vanishing Majorana spinor for supersymmetry and, in fact, a connected $7$d manifold admits a G$_2$-structure if and only if it is both spin and orientable \cite{MR2282011}. In the simplest case this structure will be torsion-free leading to a Ricci-flat manifold with holonomy group G$_2$, but one should in general anticipate flux backgrounds where this is not the case \cite{Hull:1986kz, Strominger:1986uh, Gauntlett:2001ur, Friedrich:2001nh, Friedrich:2001yp, Gauntlett:2003cy}.}  can be characterised from the world-sheet viewpoint by having an extended chiral symmetry
\be
({\cal N}=1)_{c=\frac{21}{2}} \subset {\cal SW}_{[0,\frac{21}{2}]}(\tfrac{3}{2},\tfrac{3}{2},2) \ . \label{SV}
\ee
Here ${\cal SW}_{[0,\frac{21}{2}]}(\tfrac{3}{2},\tfrac{3}{2},2)$ is a ${\cal W}$-algebra that contains an ${\cal N}=1$ superconformal subalgebra  with central charge $c=\frac{21}{2}$, as well as additional ${\cal N}=1$ multiplets of spin $(\frac{3}{2},2)$ and $(2,\frac{5}{2})$.  (The full algebra is therefore generated by three fields of spin $\frac{3}{2}$, $\frac{3}{2}$, and $2$, explaining the terminology in (\ref{SV}).) This is the G$_2$-analogue of the more familiar statement that string backgrounds of the form 
\be\label{CY}
\mathbb{M}_4 \times {\cal M}_6  \ , 
\ee
where ${\cal M}_6$ is a $6$-dimensional Calabi-Yau manifold, are characterised by having ${\cal N}=2$ superconformal symmetry on the world-sheet \cite{Banks:1987cy}.
 
One of the remarkable properties of the Shatashvili-Vafa algebra is that it actually contains, in addition to the overall ${\cal N}=1$ superconformal algebra at $c=\frac{21}{2}$, a further ${\cal N}=1$ superconformal subalgebra at $c=\frac{7}{10}$, i.e.\ a tricritical Ising model algebra \cite{Shatashvili:1994zw}. This is necessary for spacetime supersymmetry in much the same way as the $\widehat{\mathfrak{u}}(1)$ of $\mathcal{N}=2$ theories is required for the case of eq.~(\ref{CY}). It gives rise to a number of striking properties. In particular, it suggests that the algebra has special fields that behave analogously to chiral primaries of ${\cal N}=2$ superconformal field theories in that they capture some topological information about the background and describe the exactly marginal deformations. Furthermore, there were hints that one may also be able to define a topological twist for these theories, again generalising a very influential and successful construction for ${\cal N}=2$ backgrounds \cite{Shatashvili:1994zw}, see also \cite{deBoer:2005pt}. 

The Shatashvili-Vafa algebra is actually a particular element of a $2$-parameter family of ${\cal W}$-algebras ${\cal SW}_{[\lambda^2,c]}(\tfrac{3}{2},\tfrac{3}{2},2)$ first constructed in \cite{Blumenhagen:1991nm}, see also \cite{Blumenhagen:1992vr,Figueroa-OFarrill:1996tnk}. These algebras all have the same spin spectrum, and are characterised by a coupling constant $\lambda^2$ as well as the central charge, and the Shatashvili-Vafa algebra is the algebra with $[\lambda^2,c] = [0,\frac{21}{2}]$.

\smallskip

In this paper we want to study type II backgrounds\footnote{We expect that a similar construction should also work for heterotic backgrounds; for the case of (\ref{flat}) the heterotic world-sheet was studied in \cite{Melnikov:2017yvz} where the Shatashvili-Vafa algebra was derived following \cite{Banks:1987cy}.} where we replace the Minkowski factor $\mathbb{M}_3$ in (\ref{flat}) by an ${\rm AdS}_3$ factor with pure NS-NS flux; the remaining ${\cal M}_7$  manifold then also has to have some flux (to compensate for the flux through ${\rm AdS}_3$).  
We want to ask whether for backgrounds with minimal spacetime supersymmetry, there is a generalisation of the Shatashvili-Vafa algebra,\footnote{The existence of such an algebra was already forseen in \cite{Giveon:1999jg}, and some more details were surmised as part of a study of non-critical strings into singular Spin$(7)$ holonomy target spaces \cite{Sriharsha:2006zc}.}
and if so, what properties it possesses. As we shall see there is a natural candidate, namely a certain $1$-parameter family of algebras of the form ${\cal SW}_{[\lambda^2,c]}(\tfrac{3}{2},\tfrac{3}{2},2)$ that we shall denote by 
\be
({\cal N}=1)_{c=\frac{21}{2} - \frac{6}{k}} \subset {\cal SW}_{k}(\tfrac{3}{2},\tfrac{3}{2},2) \ , 
\ee
where $k>0$ labels the flux through the ${\rm AdS}_3$ factor (that needs to be compensated by the ${\cal M}_7$ manifold). We motivate our proposal by studying backgrounds of the form 
\be\label{3cases}
{\rm AdS}_3 \times {\rm S}^3 \times \mathbb{T}^4 \ , \qquad 
{\rm AdS}_3 \times {\rm S}^3 \times \text{K3} \ , \qquad 
{\rm AdS}_3 \times \bigl( {\rm S}^3 \times \mathbb{T}^4 \bigr) / D_n \ , 
\ee
all of which possess this ${\cal SW}_{k}(\tfrac{3}{2},\tfrac{3}{2},2)$ symmetry. Quite remarkably, the resulting algebra always (i.e.\ for all values of $k$) contains a tricritical Ising model with $c=\frac{7}{10}$ as a subalgebra. Furthermore, in the limit $k\rightarrow \infty$ for which ${\rm AdS}_3$ approximates flat space, it reduces to the Shatashvili-Vafa algebra. 

We then go on to study the representation theory of ${\cal SW}_{k}(\tfrac{3}{2},\tfrac{3}{2},2)$ from first principles; while some of this had already been analysed in \cite{Noyvert:2002mc}, we give a somewhat more systematic description that is also applicable in the regime where $c=\frac{21}{2} - \frac{6}{k} \geq \frac{5}{2}$. (The analysis of \cite{Noyvert:2002mc} was mainly concerned with the complementary regime $c<\frac{5}{2}$, where the algebra can be rational.) While there are again certain primaries that behave similarly to what was observed in \cite{Shatashvili:1994zw}, some of the special properties, in particular regarding the connection between the topological Betti numbers and exactly marginal operators, do not seem to hold for general $k$. We also study the analogue of the topological twist, in particular following \cite{deBoer:2005pt}. However, at least the most direct implementation of the definition of the BRST operator of \cite{deBoer:2005pt} seems to have problems, and they even persist in the flat limit, i.e.\ the original Shatashvili-Vafa set-up, see Sections \ref{sec:nilpotent} and \ref{sec:coho}. 
\medskip

The paper is organised as follows. In Section~\ref{sec:T4} we study backgrounds of the form ${\rm AdS}_3 \times {\rm S}^3 \times {\cal M}_4$ and show that the world-sheet algebra contains a specific $1$-parameter family of ${\cal SW}_{[\lambda^2,c]}(\frac{3}{2},\frac{3}{2},2)$ algebras that we identify, see eq.~(\ref{eq:c&lambda2}). (The analysis for the third case in eq.~(\ref{3cases}) is explained in Appendix~\ref{app:Dn}.) The algebra is only consistent provided that a certain null-vector $N$, see eq.~(\ref{eq:N}), is set to zero, and this allows us to give a transparent and systematic analysis of the representation theory; this is done in Section~\ref{sec:Reps}. In Section~\ref{sec:impl} we study various properties of the resulting structure: in particular, we identify the Ramond ground states and show that their fusion defines special NS primaries saturating a unitarity bound, see Section~\ref{sec:notsospecial}. However, unlike the situation in \cite{Shatashvili:1994zw}, these NS primaries are not directly related to exactly marginal operators. The latter are found in Section~\ref{sec:chiralp}, where we also define a sense in which they can be viewed as `chiral primaries'. We also study to which extent one can define a cohomology operator for these theories, see Section~\ref{sec:nilpotent}, and revisit the analysis of the cohomology of \cite{deBoer:2005pt} in the flat space limit in Section~\ref{sec:coho}. Section~\ref{sec:discussion} contains a discussion of our results, and some of the more technical aspects of our analysis are spelled out in a number of appendices.

\section{Deformed Shatashvili-Vafa algebra from $\text{AdS}_3\times\text{S}^3\times {\cal M}_4$} \label{sec:T4}

In this section we describe the analogue of the Shatashvili-Vafa algebra \cite{Shatashvili:1994zw} for the situation where the background is
\be
{\rm AdS}_3 \times {\rm S}^3 \times {\cal M}_4 \ , 
\ee
with ${\cal M}_4$ hyper-K\"ahler (i.e.\ either ${\cal M}_4=\mathbb{T}^4$ or ${\cal M}_4={\rm K3}$). A similar analysis also works for the background ${\rm AdS}_3 \times ({\rm S}^3 \times \mathbb{T}^4)/D_n$ with ${\cal N}=(2,2)$ spacetime supersymmetry \cite{Datta:2017ert}; this is described in Appendix~\ref{app:Dn}.

Assuming that the flux of the background is pure NS-NS, the world-sheet algebra corresponding to the seven-dimensional ${\rm S}^3 \times {\cal M}_4$ part of the background is\footnote{In order for this to lead to a critical string background, the level of the $\widehat{\mathfrak{sl}}(2,\mathds{R})_k^{(1)}$ algebra, describing the ${\rm AdS}_3$ factor, is  also $k$; this level is related to the ${\rm AdS}_3$ radius $R_{\rm AdS}$ as $k = R_{\rm AdS}^2 / \alpha'$.} \cite{Giveon:1998ns}
\be\label{AdS3alg}
\widehat{\mathfrak{su}}(2)_k^{(1)} \oplus ({\cal N}=4)_{c=6} \ \cong  \ 
\widehat{\mathfrak{su}}(2)_{k-2} \oplus  \text{Fer}^3 \oplus ({\cal N}=4)_{c=6} \ , 
\ee
where we have, as usual, rewritten the ${\cal N}=1$ superaffine algebra $\widehat{\mathfrak{su}}(2)_k^{(1)}$ in terms of the decoupled bosonic currents generating $\widehat{\mathfrak{su}}(2)_{k-2}$ (denoted by $\mathcal{K}^i$ with $i=1,2,3$), and three free fermions ($\chi^i$ with $i =1, 2, 3$); more details about our conventions are spelled out in Appendix~\ref{sec:su(2)conventions}. The ${\cal N}=4$ algebra at $c=6$ arises from ${\cal M}_4$, and the currents of the $R$-symmetry of this ${\cal N}=4$ algebra are denoted by $\jmath^i$ with OPE 
\begin{equation} \label{eq:su(2)1forK3}
\jmath^i(z)\jmath^j(w) \ \sim \ \frac{\delta^{ij}}{2(z-w)^2} + \frac{i\epsilon^{ij}{}_k \,\jmath^k(w)}{(z-w)}
\ .
\end{equation}
They generate an $\widehat{\mathfrak{su}}(2)_1$ algebra (since the ${\cal N}=4$ algebra has central charge $c=6$). 

In order to motivate our ansatz for the analogue of the Shatashvili-Vafa  algebra, let us denote by $\{\omega^i\}_{i=1,\,2,\,3}$ the hyper-K\"ahler triplet of closed 2-forms on ${\cal M}_4$. Since S$^3=\text{SU}(2)$ is parallelisable, we can 
choose a global coframe $\{e^i\}_{i=1,\,2,\,3}$ on the $3$-sphere, and specify a G$_2$-structure on S$^3\times {\cal M}_4$ (with the product metric) by defining the G$_2$-invariant 3-form
\begin{equation}
\varphi = \sum_{i=1}^3 e^i\wedge \omega^i - e^1\wedge e^2\wedge e^3 \ .
\end{equation}
In order to mimic this on the world-sheet we identify\footnote{
These identifications are motivated by semi-classical chiral symmetries, see \cite{delaOssa:2018azc, Howe:1991ic}.} $e^i$ with the fermions $\chi^i$ coming from the S$^3$, and $\omega^i$ with the currents $\jmath^i$ generating the $R$-symmetry. Thus the world-sheet field corresponding to $\varphi$ is taken as
\begin{equation} \label{eq:PforK3}
P=-i\sqrt{\frac{8}{k}}\sum_{i=1}^3 \chi^i\jmath^i + \sqrt{\frac{8}{k^3}}\, \chi^1\chi^2\chi^3 \ , 
\end{equation}
where the overall normalisation has been chosen for future convenience. We also have an ${\cal N}=1$ superconformal algebra generated by 
\begin{align}
T&=T^{\text{S}^3}+T^{(\mathcal{N}=4)} \ ,  \label{N1L} \\
G&=G^{\text{S}^3}+G^{(\mathcal{N}=4)}\ , \label{N1G}
\end{align}
where $T^{\text{S}^3}$ and $G^{\text{S}^3}$ are the ${\cal N}=1$ fields of eq.~\eqref{eq:TGS3} associated to the superaffine algebra $\widehat{\mathfrak{su}}(2)_k^{(1)}$, while $T^{(\mathcal{N}=4)}$ and $G^{(\mathcal{N}=4)}=\frac{1}{\sqrt{2}}(G^+ + G^-)$ are the ${\cal N}=1$ fields coming from the ${\cal N}=4$ factor. 

The idea of the construction is to determine the chiral algebra that is generated by the fields $P$, $T$ and $G$.  First of all, the OPE of $P$ with $P$ leads to 
\be
P(z)\, P(w) \ \sim \ \frac{-7}{(z-w)^3} + \frac{6 X(w)}{(z-w)} \ , 
\ee
where $X$ is the field 
\begin{equation}
X=-\sum_{i=1}^3\left(\frac{2}{3}\jmath^i\jmath^i+\frac{1}{k}\partial\chi^i\chi^i\right)-\frac{4i}{k}\left(\chi^1\chi^2\jmath^3-\chi^1\chi^3\jmath^2+\chi^2\chi^3\jmath^1\right)\,.
\end{equation}
We mention in passing that, just as in \cite{Shatashvili:1994zw}, this current corresponds essentially to the Hodge dual $4$-form of the G$_2$ $3$-form $\varphi$ in the $7$-dimensional target space (both of which are fixed by the G$_2$ action).
Furthermore, under the action of the ${\cal N}=1$ superalgebra, $P$ and $X$ generate two supermultiplets: the bosonic partner of  $P$ is $K=G_{-1/2} P$ of spin $h=2$, while the fermionic partner of  $X$ is $M=G_{-1/2} X$ of spin $h=5/2$; thus the total spectrum is 
\be
\underbrace{(G,T)}_{(\frac{3}{2},2)}  \ \qquad 
\underbrace{(P,K)}_{(\frac{3}{2},2)}  \ \qquad 
\underbrace{(X,M)}_{(2,\frac{5}{2})}  \ .
\ee
We have checked that these fields close among themselves (up to normal ordered products), and the complete set of OPEs is spelled out in Appendix~\ref{app:SW_kOPEs}. (Note that the closure of the algebra uses the fact that $\widehat{\mathfrak{su}}(2)_1$ has null-vectors at level $2$.)

The complete algebra is thus generated by three ${\cal N}=1$ multiplets of spin $\frac{3}{2}$, $\frac{3}{2}$, and $2$, and it therefore describes a ${\cal W}$-algebra of type $\mathcal{SW}(\frac{3}{2},\frac{3}{2},2)$ in the nomenclature of \cite{Bouwknegt:1992wg}. There is a 2-parameter family of such algebras, as originally discovered by Blumenhagen \cite{Blumenhagen:1991nm}. One parameter is the central charge $c$ and following \cite{Noyvert:2002mc} the second parameter will be denoted by  $\lambda^2$. Here $\lambda$ denotes the self-coupling constant of the spin $\frac{3}{2}$ multiplet $(P,K)$, see eq.~(2.3) of \cite{Noyvert:2002mc}, and it follows from \cite[Appendix~A]{Noyvert:2002mc} that, up to a rescaling of the generators, the OPEs only depend on $\lambda^2$ and $\mu^2 = \frac{9c(4+\lambda^2)}{2(27-2c)}$.\footnote{Thus the algebra is really characterised by $\lambda^2$ (and $c$) rather than $\lambda$ (and $c$).}

From the explicit OPEs of Appendix~\ref{app:SW_kOPEs} we find that the parameters take the values 
\begin{equation} \label{eq:c&lambda2}
c=\frac{21}{2}-\frac{6}{k}
\qquad\quad\text{and}\qquad\quad
\lambda^2=\frac{32(3k-2)^2}{k^2 (49k-30)}\ .
\end{equation}
Our ${\cal W}$-algebras therefore define a $1$-parameter family of $\mathcal{SW}(\frac{3}{2},\frac{3}{2},2)$ algebras, which we shall denote by $\mathcal{SW}_k(\frac{3}{2},\frac{3}{2},2)$  in the following.  Note that the family (\ref{eq:c&lambda2}) can be more abstractly characterised by the property that 
\begin{equation}\label{lamc}
\lambda^2 = -\frac{4 (3-2 c)^2 (2 c-21)}{27 (10 c-7)}\ .
\end{equation}
\smallskip

In the general case, the algebra $\mathcal{SW}(\frac{3}{2},\frac{3}{2},2)$  has a coset realisation as \cite{Noyvert:2002mc}
\be\label{coset}
{\cal SW}(\tfrac{3}{2},\tfrac{3}{2},2) = \frac{\widehat{\mathfrak{su}}(2)_{k_1} \oplus \widehat{\mathfrak{su}}(2)_{k_2} \oplus  \widehat{\mathfrak{su}}(2)_{2}}{\widehat{\mathfrak{su}}(2)_{k_1+k_2+2}} \ , 
\ee
where\footnote{Strictly speaking, $\text{Fer}^3$ is isomorphic to $\widehat{\mathfrak{su}}(2)_{2}$, together with the $j=1$ representation of $\widehat{\mathfrak{su}}(2)_{2}$, which has $h=\frac{1}{2}$; the three lowest weight states in the $j=1$ representation of $\widehat{\mathfrak{su}}(2)_{2}$ are the three free fermions themselves.} $\widehat{\mathfrak{su}}(2)_{2} \cong \text{Fer}^3$
and $c$ and $\lambda^2$ can be expressed in terms of $k_1$ and $k_2$ as 
\begin{eqnarray} 
c & = &  \frac{3k_1}{k_1+2} + \frac{3k_2}{k_2+2} + \frac{3}{2} - \frac{3(k_1+k_2+2)}{k_1+k_2+4} \ , 
\label{eq:cNoyvert} \\ 
\lambda^2 & = & \frac{32 (k_1-k_2)^2 (2 k_1+k_2+6)^2 (k_1+2 k_2+6)^2}{27 k_1k_2  (k_1+2)^2 (k_2+2)^2 (k_1+k_2+4)^2 (k_1+k_2+6)} \ . \label{eq:lambda2Noyvert}
\end{eqnarray}
Given that we can (formally) write this coset in either of the following forms 
\begin{eqnarray}
{\cal SW}(\tfrac{3}{2},\tfrac{3}{2},2) & = & \boxed{\frac{\widehat{\mathfrak{su}}(2)_{k_1} \oplus   \widehat{\mathfrak{su}}(2)_{2}}{\widehat{\mathfrak{su}}(2)_{k_1+2}}} \oplus \frac{\widehat{\mathfrak{su}}(2)_{k_1+2} \oplus  \widehat{\mathfrak{su}}(2)_{k_2}}{\widehat{\mathfrak{su}}(2)_{k_1+k_2+2}}  \\[4pt]
& = & \boxed{\frac{\widehat{\mathfrak{su}}(2)_{k_2} \oplus   \widehat{\mathfrak{su}}(2)_{2}}{\widehat{\mathfrak{su}}(2)_{k_2+2}}} \oplus \frac{\widehat{\mathfrak{su}}(2)_{k_2+2} \oplus  \widehat{\mathfrak{su}}(2)_{k_1}}{\widehat{\mathfrak{su}}(2)_{k_1+k_2+2}}  \\[4pt]
& = & \frac{\widehat{\mathfrak{su}}(2)_{k_1} \oplus   \widehat{\mathfrak{su}}(2)_{k_2}}{\widehat{\mathfrak{su}}(2)_{k_1+k_2}} \oplus \boxed{ \frac{\widehat{\mathfrak{su}}(2)_{k_1+k_2} \oplus  \widehat{\mathfrak{su}}(2)_{2}}{\widehat{\mathfrak{su}}(2)_{k_1+k_2+2}}  }
\end{eqnarray}
together with the fact that 
\be
\frac{\widehat{\mathfrak{su}}(2)_n \oplus \widehat{\mathfrak{su}}(2)_2}{\widehat{\mathfrak{su}}(2)_{n+2}} \cong \bigl( {\cal N}=1 \bigr)_{c_n} \ \ \hbox{with} \ \ c_n = \frac{3}{2} - \frac{12}{(n+2)(n+4)} \ , 
\ee
it follows that ${\cal SW}(\tfrac{3}{2},\tfrac{3}{2},2)$ generically contains $3$ distinct ${\cal N}=1$ subalgebras with central charges 
\be\label{ck}
c_{k_1} \ , \qquad c_{k_2} \ , \qquad c_{k_1 + k_2} \ ,
\ee
in addition to the overall one with central charge $c$ of eq.~(\ref{eq:cNoyvert}).
Recall that one of the special properties of the Shatashvili-Vafa algebra \cite{Shatashvili:1994zw}  is that it contains an ${\cal N}=1$ superconformal subalgebra with $c_1=\frac{7}{10}$, i.e.\ a tricritical Ising model. It turns out that this property also continues to hold for the deformed algebra $\mathcal{SW}_k(\frac{3}{2},\frac{3}{2},2)$ we have constructed above, i.e.\ the algebra generated by $P$, $G$ and $T$. Indeed, the relevant generators are simply 
\begin{equation} \label{eq:TricriticalGenerators}
\tilde{T} = -\frac{1}{5}X 
\qquad\quad\text{and}\quad\qquad
\tilde{G} = \frac{i}{\sqrt{15}}P\ .
\end{equation}
This can also be understood in terms of the coset construction of eq.~(\ref{coset}): the deformed Shatashvili-Vafa algebra $\mathcal{SW}_k(\frac{3}{2},\frac{3}{2},2)$ corresponds to either $k_1=1$, or $k_2=1$, or $k_1+k_2=-7$,\footnote{Other branches containing a tricritical Ising model are specified by $k_1 = -7$, or $k_2=-7$, or $k_1+k_2=1$, but they do not satisfy eq.~(\ref{lamc}).} and in each case, one of the three ${\cal N}=1$ algebras in (\ref{ck}) has central charge $\frac{7}{10}$.  Obviously, for the branch $(k_1,k_2)=(1,k_2)$, say, we can express $k$ in terms of $k_2$ and vice versa, but the explicit form of the relation is not particularly simple nor illuminating. 

In the limit $k\rightarrow \infty$ both AdS$_3$ and S$^3$ have infinite radius and are thus flat. In this limit, 
\begin{equation}
c=\frac{21}{2}
\qquad\quad\text{and}\qquad\quad
\lambda^2=0\,,
\end{equation}
which is precisely the usual flat space Shatashvili-Vafa algebra \cite{Noyvert:2002mc}. This is nicely consistent with our algebra being a finite $k$ deformation thereof. This can also be seen from the explicit OPEs of Appendix~\ref{app:SW_kOPEs}  that reproduce the OPEs of the Shatashvili-Vafa algebra, see e.g.\ \cite[App.~A.5]{Fiset:2018huv}, up to $i\sqrt{1/k}$ corrections. To see this one also needs to take into account that, just as the usual Shatashvili-Vafa algebra \cite{Figueroa-OFarrill:1996tnk}, its deformation is only associative up to null fields;\footnote{Indeed, there is a small mismatch in the order one pole of the $M M$ OPE, which is independent of $k$. We have verified that, up to adding null fields, it is proportional to $\sqrt{1/k}$ and hence vanishes in the limit.} we elaborate on this point in Section~\ref{sec:Associativity} below.

Finally, the coset construction implies that there is yet another way in which we may realise a ${\cal SW}(\tfrac{3}{2},\tfrac{3}{2},2)$ algebra in the world-sheet algebra of eq.~(\ref{AdS3alg}): since the ${\cal N}=4$ algebra contains an $\widehat{\mathfrak{su}}(2)$ subalgebra at level $\frac{c}{6}=1$, the world-sheet algebra actually contains 
\be\label{addc}
\widehat{\mathfrak{su}}(2)_{k-2} \oplus  \widehat{\mathfrak{su}}(2)_{2} \oplus 
\widehat{\mathfrak{su}}(2)_{1} \subset 
\widehat{\mathfrak{su}}(2)_k^{(1)} \oplus ({\cal N}=4)_{c=6} \ , 
\ee
where we have again used that $ \text{Fer}^3 \cong \widehat{\mathfrak{su}}(2)_{2}$. This leads to an ${\cal SW}(\tfrac{3}{2},\tfrac{3}{2},2)$ algebra with $(k_1, k_2)=(1,k-2)$, which therefore satisfies \eqref{lamc}. However, the overall ${\cal N}=1$ superconformal algebra of this ${\cal SW}(\tfrac{3}{2},\tfrac{3}{2},2)$ does not agree with that of ${\rm S}^3 \times {\cal M}_4$ since 
its central charge is smaller than $\frac{21}{2} - \frac{6}{k}$, see eq.~(\ref{eq:c&lambda2}), which is the central charge of the right-hand-side of eq.~(\ref{addc}). Indeed,
the central charge of $\widehat{\mathfrak{su}}(2)_1$ is $c=1$, which is smaller than that of the ${\cal N}=4$ algebra (which is $c=6$), and the quotient by $\widehat{\mathfrak{su}}(2)_{k+1}$ in the coset construction of eq.~(\ref{coset}) reduces the central charge of the coset further. 

\subsection{Two different regimes} \label{sec:two}

For the application we have in mind where at least typically $k\geq 1$ --- we will comment on this bound in more detail momentarily, see eq.~(\ref{kbound}) below --- the central charge of eq.~(\ref{eq:c&lambda2}) satisfies $c\geq \frac{9}{2}$. On the other hand, $c=\frac{9}{2}$ is the largest central charge that can be reached by the rational coset CFTs  $\mathcal{SW}(\frac{3}{2},\frac{3}{2}, 2)$ in (\ref{coset}),  i.e.\ for the case where both $k_1$ and $k_2$ are positive integers. Thus we will be mainly interested in the non-rational regime. 

To be a bit more specific, we are interested in the algebra $\mathcal{SW}_k(\frac{3}{2},\frac{3}{2}, 2)$ corresponding to the coset branch with $k_1=1$, say. Then the rational models appear for $c< \frac{5}{2}$, which corresponds to $k < \frac{3}{4}$ in eq.~(\ref{eq:c&lambda2}). Thus provided that 
\be\label{kbound}
k\geq \frac{3}{4} \quad \Longrightarrow \quad \hbox{$\mathcal{SW}_k(\frac{3}{2},\frac{3}{2}, 2)$ is non-rational.}
\ee
If $k<\frac{3}{4}$, the algebra can only be unitary provided that all three ${\cal N}=1$ subalgebras in (\ref{ck}) are unitary, which precisely matches the condition that $k_2$ is a positive integer for $k_1=1$. (In this case, all three of these subalgebras are ${\cal N}=1$ minimal models.) This is the regime that was studied in detail in \cite{Noyvert:2002mc}.

\begin{figure}
\begin{center}
\begin{tikzpicture}{scale=1}
\draw[->] (-0.5, 0) -- (11, 0) node[right] {$c$};
\draw[->] (0, -0.5) -- (0, 2.5) node[above] {$\lambda^2$};

\pgfmathsetmacro\k{1}
\draw[scale=1, domain={3*(3*\k+2)/2/(\k+2)}:{9*(6+\k)/2/(\k+2)}, variable=\c, red, thick] plot ({\c},{4*(4*\c-9*\k+2*\c*\k)/243*(4*\c-9*\k+2*\c*\k)/\k/(16*\c-18*\k+12*\c*\k-3*\k^2+2*\c*\k^2)*(54-4*\c+9*\k-2*\c*\k)});
\draw ({9*(6+\k)/2/(\k+2)},0.1) -- ({9*(6+\k)/2/(\k+2)},-0.1) node[below] {\footnotesize{$\frac{21}{2}$}};
\draw ({9*\k/(4+2*\k)},0.1) -- ({9*\k/(4+2*\k)},-0.1) node[below] {\footnotesize{$\frac{3}{2}$}};

\pgfmathsetmacro\k{2}
\draw[scale=1, domain={3*(3*\k+2)/2/(\k+2)}:{9*(6+\k)/2/(\k+2)}, variable=\c, black, thick] plot ({\c},{4*(4*\c-9*\k+2*\c*\k)/243*(4*\c-9*\k+2*\c*\k)/\k/(16*\c-18*\k+12*\c*\k-3*\k^2+2*\c*\k^2)*(54-4*\c+9*\k-2*\c*\k)});
\draw ({9*(6+\k)/2/(\k+2)},0.1) -- ({9*(6+\k)/2/(\k+2)},-0.1) node[below] {\footnotesize{$9$}};

\pgfmathsetmacro\k{3}
\draw[scale=1, domain={3*(3*\k+2)/2/(\k+2)}:{9*(6+\k)/2/(\k+2)}, variable=\c, black, thick] plot ({\c},{4*(4*\c-9*\k+2*\c*\k)/243*(4*\c-9*\k+2*\c*\k)/\k/(16*\c-18*\k+12*\c*\k-3*\k^2+2*\c*\k^2)*(54-4*\c+9*\k-2*\c*\k)});
\draw ({9*(6+\k)/2/(\k+2)},0.1) -- ({9*(6+\k)/2/(\k+2)},-0.1) node[below, fill=white] {\footnotesize{$\frac{81}{10}$}};

\pgfmathsetmacro\k{4}
\draw[scale=1, domain={3*(3*\k+2)/2/(\k+2)}:{9*(6+\k)/2/(\k+2)}, variable=\c, black, thick] plot ({\c},{4*(4*\c-9*\k+2*\c*\k)/243*(4*\c-9*\k+2*\c*\k)/\k/(16*\c-18*\k+12*\c*\k-3*\k^2+2*\c*\k^2)*(54-4*\c+9*\k-2*\c*\k)});
\draw ({9*(6+\k)/2/(\k+2)},0.1) -- ({9*(6+\k)/2/(\k+2)},-0.1) node[below, fill=white] {\footnotesize{$\frac{15}{2}$}};

\pgfmathsetmacro\k{5}
\draw[scale=1, domain={3*(3*\k+2)/2/(\k+2)}:{9*(6+\k)/2/(\k+2)}, variable=\c, black, thick] plot ({\c},{4*(4*\c-9*\k+2*\c*\k)/243*(4*\c-9*\k+2*\c*\k)/\k/(16*\c-18*\k+12*\c*\k-3*\k^2+2*\c*\k^2)*(54-4*\c+9*\k-2*\c*\k)});

\pgfmathsetmacro\k{6}
\draw[scale=1, domain={3*(3*\k+2)/2/(\k+2)}:{9*(6+\k)/2/(\k+2)}, variable=\c, black, thick] plot ({\c},{4*(4*\c-9*\k+2*\c*\k)/243*(4*\c-9*\k+2*\c*\k)/\k/(16*\c-18*\k+12*\c*\k-3*\k^2+2*\c*\k^2)*(54-4*\c+9*\k-2*\c*\k)});

\pgfmathsetmacro\k{7}
\draw[scale=1, domain={3*(3*\k+2)/2/(\k+2)}:{9*(6+\k)/2/(\k+2)}, variable=\c, black, thick] plot ({\c},{4*(4*\c-9*\k+2*\c*\k)/243*(4*\c-9*\k+2*\c*\k)/\k/(16*\c-18*\k+12*\c*\k-3*\k^2+2*\c*\k^2)*(54-4*\c+9*\k-2*\c*\k)});

\pgfmathsetmacro\k{8}
\draw[scale=1, domain={3*(3*\k+2)/2/(\k+2)}:{9*(6+\k)/2/(\k+2)}, variable=\c, black, thick] plot ({\c},{4*(4*\c-9*\k+2*\c*\k)/243*(4*\c-9*\k+2*\c*\k)/\k/(16*\c-18*\k+12*\c*\k-3*\k^2+2*\c*\k^2)*(54-4*\c+9*\k-2*\c*\k)});

\pgfmathsetmacro\k{9}
\draw[scale=1, domain={3*(3*\k+2)/2/(\k+2)}:{9*(6+\k)/2/(\k+2)}, variable=\c, black, thick] plot ({\c},{4*(4*\c-9*\k+2*\c*\k)/243*(4*\c-9*\k+2*\c*\k)/\k/(16*\c-18*\k+12*\c*\k-3*\k^2+2*\c*\k^2)*(54-4*\c+9*\k-2*\c*\k)});

\pgfmathsetmacro\k{10}
\draw[scale=1, domain={3*(3*\k+2)/2/(\k+2)}:{9*(6+\k)/2/(\k+2)}, variable=\c, black, thick] plot ({\c},{4*(4*\c-9*\k+2*\c*\k)/243*(4*\c-9*\k+2*\c*\k)/\k/(16*\c-18*\k+12*\c*\k-3*\k^2+2*\c*\k^2)*(54-4*\c+9*\k-2*\c*\k)});

\foreach \l in {1,...,25}{
\node at ({3*1/(1+2)+3*\l/(\l+2)+3/2-3*(1+\l+2)/(1+\l+4)},{32/1/\l/(1+2)^2/27*(1-\l)^2*(2*1+\l+6)^2/(1+\l+4)^2/(1+\l+6)*(1+2*\l+6)^2/(\l+2)^2}) {\Large \textcolor{red}{$\cdot$}};
}

\foreach \k in {2,...,3}{
\foreach \l in {\k,...,25}{
\node at ({3*\k/(\k+2)+3*\l/(\l+2)+3/2-3*(\k+\l+2)/(\k+\l+4)},{32/\k/\l/(\k+2)^2/27*(\k-\l)^2*(2*\k+\l+6)^2/(\k+\l+4)^2/(\k+\l+6)*(\k+2*\l+6)^2/(\l+2)^2}) {\Large $\cdot$};
}}

\draw[scale=1, domain=1.88:4.5, variable=\c, blue, line width=2pt, samples=1000] plot ({\c},{4*(9-2*\c)^2/243/(2*\c-3)*(9-2*\c)});
\draw (4.5,0.1) -- (4.5,-0.1) node[below] {\footnotesize{$\frac{9}{2}$}};
\draw[line width=2pt, blue] (4.5,0.1) -- (4.5,-0.1);

\draw (2.5,0.45) -- (2.5,-0.1) node[below] {\footnotesize{$\frac{5}{2}$}};
\end{tikzpicture}
\caption{The parameter values where $\mathcal{SW}_k(\frac{3}{2},\frac{3}{2}, 2)$ (red) and 
$\mathcal{SW}(\frac{3}{2},\frac{3}{2}, 2)$ (black) are unitary. The blue line discriminates between the rational regime and the continuous regime. The first three black lines correspond to the values $k_1=2$, $k_1=3$, and $k_1=4$, where the limiting values of $c$ are $c=9$, $c=\frac{81}{10}$ and $c=\frac{15}{2}$, respectively. We have only analysed the regime where $\lambda^2\geq 0$ and $0<c<\frac{21}{2}$. }
\label{fig:UnitarityLoci}
\end{center}
\end{figure}
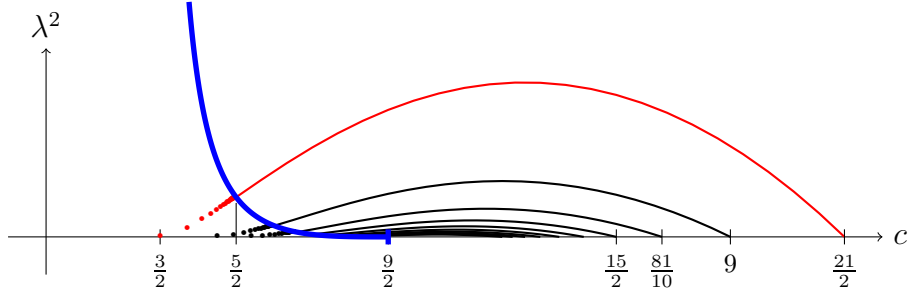

On the other hand, for $k\geq \frac{3}{4}$, it is less clear whether the theory will be unitary or not, but one may suspect that this will be generically the case. (This regime should behave similarly to that of an ${\cal N}=1$ superconformal algebra with $c\geq \frac{3}{2}$.) Assuming that all of these non-rational theories are in fact unitary, we have plotted the unitarity regime for the algebra $\mathcal{SW}_k(\frac{3}{2},\frac{3}{2}, 2)$ (red) in Figure~\ref{fig:UnitarityLoci}. The other curves (black) correspond to $k_1=2, 3,\ldots$ and complete the picture of the unitarity regime for $\mathcal{SW}(\frac{3}{2},\frac{3}{2}, 2)$.

It is intriguing that the limiting value of the central charge ($c=9$) for $k_2=2$ could potentially balance an ${\rm AdS}_4$ factor, and similarly for ${\rm AdS}_5$ ($c=\frac{15}{2}$ with $k_2=4$) and ${\rm AdS}_6$ ($c=6$ with $k_2=10$). Note that this coincidence with the target space dimensions is essentially the same as in \cite{Fiset:2020lmg}.

\subsection{Associativity} \label{sec:Associativity}

For the analysis of the representation theory it will be important that ${\cal SW}_k(\tfrac{3}{2},\tfrac{3}{2},2)$ is only consistent (i.e.\ satisfies associativity) provided that we set to zero the null-field 
\begin{equation} \label{eq:N}
N=4GX-\frac{8i}{7} \sqrt{\frac{2}{k}} PX-2PK-\partial^2G+ \frac{9i}{7} \sqrt{\frac{2}{k}} \partial^2P-4 \partial M \ .
\end{equation}
Since the large $k$ limit of $N$ is non-trivial this reflects a similar phenomenon for the Shatashvili-Vafa algebra, as was already pointed out some time ago in \cite{Figueroa-OFarrill:1996tnk}. We have checked that $N$ generates all null-vectors of the vacuum representation up to level $11/2$ for generic $k$. We should also mention that in the realisation \eqref{AdS3alg} of ${\cal SW}_k(\tfrac{3}{2},\tfrac{3}{2},2)$ corresponding to $\text{S}^3\times {\cal M}_4$, the field $N$ vanishes identically for the case of $\mathcal{M}_4 = \mathbb{T}^4$, while it is a consequence of ${\cal N}=4$ $c=6$ null-vectors in general. 

\section{Representation theory}\label{sec:Reps}

The presence of the null-vector $N$ in eq.~(\ref{eq:N}) restricts the possible representations of 
${\cal SW}_k(\tfrac{3}{2},\tfrac{3}{2},2)$ significantly, as we shall now explain. For the undeformed algebra with $[c,\lambda^2]=[\frac{21}{2},0]$ this was already used in \cite[Section 7]{Noyvert:2002mc}; here we generalise this analysis for the general deformed case.

We shall perform the analysis separately for the NS- and the R-sector.\footnote{For $\lambda^2=0$, there also exist  first and second twisted sectors \cite{Noyvert:2002mc}, but they do not generalise to the deformed set-up.} As usual fermionic generators are half-integer moded in the NS sector and integer moded in the Ramond sector; bosonic generators, on the other hand, are integer moded in both of these sectors.

We shall be interested in highest weight representations, i.e.\ representations generated from a `highest weight state' (that is annihilated by the positive modes) by the action of the negative (and zero) modes. In either case we will use the fact that the zero modes of null states must annihilate these highest weight states; the idea of characterising highest weight representations in this manner goes back to \cite{Eholzer:1991zd}, and is known as the Zhu's algebra in the mathematical literature \cite{Zhu96}, see also \cite{Gaberdiel:1998fs,Brungs:1998ij}.

\subsection{The Neveu-Schwarz sector} \label{sec:NS}

In the NS sector only the bosonic fields have zero modes, and we therefore have, in particular, zero modes for the bosonic generating fields $L_0$, $X_0$ and $K_0$. Their commutators all vanish identically, except for $[X_0, K_0]$, see Appendix~\ref{app:SW_kOPEs}. However, here we are only interested in the action of these modes on highest weight states, and we can therefore evaluate the normal product that appears on the right-hand-side of (\ref{XK}) by dropping any normal ordered term that involves positive modes. Using that 
\begin{equation}
(GP)_0 \approx -P_{1/2}G_{-1/2} \approx -\{G_{-1/2}, P_{1/2}\} = -K_0\ , 
\end{equation}
where $\approx$ denotes equality up to such positive terms, we conclude that 
\begin{equation}
[X_0, K_0] = 3K_0+3(GP)_0 \approx 0 \ .
\end{equation}
Thus we can label the highest weight states by their eigenvalues under these zero modes, 
\be
L_0 |h,\tilde{h},\kappa\rangle = h \, |h,\tilde{h},\kappa\rangle \ , \quad 
X_0 |h,\tilde{h},\kappa\rangle = - 5 \tilde{h} \, |h,\tilde{h},\kappa\rangle \ , \quad 
K_0 |h,\tilde{h},\kappa\rangle = \kappa \, |h,\tilde{h},\kappa\rangle \ ,
\ee
where $\tilde{h}$ is the eigenvalue under the tricritical Ising model stress tensor, $\tilde{T} = - \frac{1}{5} X$, see eq.~(\ref{eq:TricriticalGenerators}). 

Since the null-vector $N$ in eq.~(\ref{eq:N}) is fermionic, it does not have a zero mode in the NS sector, but we can consider instead any of its descendants, i.e.\ any linear combination of $G_{-1/2}N$, $P_{-1/2}N$, or $M_{-1/2}N$. A very stringent constraint follows from the tricritical Ising null-vector $3\partial^2 X-10\partial PP-4X^2$, which is a certain linear combination of these fields. Its zero mode equals on highest weight states
\begin{equation} \label{eq:NSfixingMMweight}
(3\partial^2X-10\partial PP-4X^2)_0 
\approx
10\tilde{L}_0(1-10\tilde{L}_0)
\overset{!}{\approx} 0 \ , 
\end{equation}
where $\tilde{L}_0 = - \frac{1}{5} X_0$, see eq.~(\ref{eq:TricriticalGenerators}). Thus we conclude that 
\be
\tilde{h} =  0 \ , \qquad \hbox{or} \qquad \tilde{h} =  \frac{1}{10} \ ,
\ee
which just reproduces the allowed NS-sector representations of the tricritical Ising model.
Similarly, the other two null-vectors constrain the representations further, and we end up with two possible cases: either we have 
\begin{flalign} \label{eq:NullConstraintNS1}
&~\underline{\hbox{Type I}_{\rm NS}:} \quad h \text{ unconstrained} \ ,\qquad
\tilde{h}=0\ ,\qquad
\kappa=0\ ,  &&
\end{flalign}
or
\begin{flalign} \label{eq:NullConstraintNS2}
&~\underline{\hbox{Type II}_{\rm NS}:} \quad h = \frac{1}{2}+\frac{\kappa}{2}\left(i\sqrt{\frac{2}{k}}-\kappa\right)\ ,\qquad
\tilde{h}=\frac{1}{10}\ ,\qquad
\kappa \text{ unconstrained.} &&
\end{flalign}
Alternatively, we can arrive at the same result by demanding that 
\begin{equation}
\begin{split}
&N_{-1/2}\ket{h, \tilde{h}, \kappa}\\
&\quad=\left(\big(1+4x\big)G_{-1/2} - \Big( 2 \kappa + \frac{4 i}{7} \sqrt{\frac{2}{k}}(1 + 2 x)\Big)P_{-1/2}-2M_{-1/2}\right)\ket{h, \tilde{h}, \kappa} \label{Nexp}
\end{split}
\end{equation}
is null, i.e.\ that it is annihilated by the positive modes $G_{1/2}$, $P_{1/2}$ and $M_{1/2}$; this leads to the constraints 
\begin{align}
G_{1/2}N_{-1/2}\ket{h, \tilde{h}, \kappa}=0 ~~ &\Rightarrow ~~
 c_2 \kappa + 2 c_1 h + c_3 \left(h- i \sqrt{\tfrac{2}{k}} \kappa - 10 \tilde{}\right) = 0 \\
P_{1/2}N_{-1/2}\ket{h, \tilde{h}, \kappa}=0 ~~ &\Rightarrow ~~
c_1 \kappa + \frac{5 c_3 \kappa}{2} - 30 c_2 \tilde{h} = 0
\\
M_{1/2}N_{-1/2}\ket{h, \tilde{h}, \kappa}=0 ~~ \nonumber &\\
 \Rightarrow ~~
10 c_1 \tilde{l} - 2 c_2 \Bigl(\kappa -&15 i \sqrt{\tfrac{2}{k}}\tilde{h}\Bigr) - c_3 \left(\kappa^2 + 2(30 l - 12 - 5 \tilde{h}) \tilde{h}\right) = 0
\ ,
\end{align}
where $c_1$, $c_2$, and $c_3$ are the coefficients of $G_{-1/2}$, $P_{-1/2}$ and $M_{-1/2}$ in (\ref{Nexp}), respectively. The solutions to this system of equations then also lead to either (\ref{eq:NullConstraintNS1}) or (\ref{eq:NullConstraintNS2}). These constraints are the finite $k$ generalisation of the constraints found in \cite{Noyvert:2002mc}.\footnote{There seems to be a factor of a half missing in \cite{Noyvert:2002mc} next to $x$ in the expression for $h$ in the last line of his equation (7.6).} Our analysis also completes the analysis of \cite{deBoer:2005pt}  (in the $k\rightarrow\infty$ limit) where the null constraints were not imposed systematically.

\subsubsection{Unitarity}

\begin{figure}[thb]
\begin{center}
\begin{tikzpicture}

\draw[->] (-5, 0) -- (3, 0) node[right] {$i\kappa$};
\draw[->] (0, -1) -- (0, 3) node[above] {$h$};

\draw[scale=2, domain={-2}:{1}, variable=\x, black, thick] plot ({\x},{1/2+\x/2*((2/3)^(1/2)+\x)});

\draw[dashed] (-1.633,1) -- (0,1) -- (2,1) node[right] {$\frac{1}{2}$};
\fill (-0.8165,0.8333) circle (0.07);
\draw[dashed] (-0.8165,0.8333) -- (0,0.8333) -- (0.5,0.5) node[right] {$\frac{1}{2}-\frac{1}{4k}$};
\draw[dashed] (-1.633,1) -- (-1.633,-0.2) node[below] {$-\sqrt{\frac{2}{k}}$};
\end{tikzpicture}
\end{center}
\caption{$h$ vs.\ $i\kappa$ for Type II$_{\rm NS}$ representation \eqref{eq:NullConstraintNS2}. The minimum is $h\geq \frac{1}{2}-\frac{1}{4k}$.}
\label{fig:L0ofK0}
\end{figure}
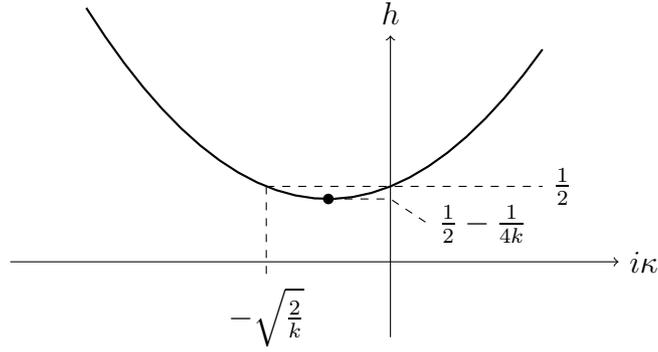

The attentive reader will have noticed that so far we have not used unitarity at all, but only the self-consistency of the chiral algebra, i.e.\ the requirement that $N=0$ has to act by zero so that the operator product is associative. In order to impose unitarity we first need to define the action of hermitian conjugation on the modes; the most natural definition that is compatible with the algebra relations of Appendix~\ref{app:SW_kOPEs} is
\begin{equation}
\begin{split}
L_n^\dagger&=L_{-n}\,,\qquad
P_r^\dagger=-P_{-r}\,,\qquad~~
X_n^\dagger=X_{-n}\,,\qquad
\\
G_r^\dagger&=G_{-r}\,,\qquad
K_n^\dagger=-K_{-n}\,,\qquad
M_r^\dagger=-M_{-r}+\frac{1}{2}G_{-r}-i\sqrt{\frac{2}{k}}P_{-r}\ . 
\end{split}
\label{conj}
\end{equation}
Note that $M$ is not quasiprimary, and hence its hermitian conjugate also involves the modes of $L_1 M = \frac{1}{2} G - i \sqrt{\frac{2}{k}} P$. One important consquence of (\ref{conj}) is that $K_0$ is anti-hermitian, and hence $\kappa$ must be purely imaginary, while $h$, being the eigenvalue of the hermitian operator $L_0$, must be real. For the representations of Type II$_{\rm NS}$, see eq.~(\ref{eq:NullConstraintNS2}),  this leads to the bound, see also Figure~\ref{fig:L0ofK0},
\begin{equation} \label{eq:NSbound110}
h \geq \frac{1}{2}-\frac{1}{4k}
\qquad\qquad (\hbox{Type II$_{\rm NS}$ representation with $\tilde{h} = \tfrac{1}{10}$})\ .
\end{equation}
On the other hand, for the representations of Type I$_{\rm NS}$, see eq.~(\ref{eq:NullConstraintNS1}), the overall Virasoro algebra relations imply that 
\begin{equation} \label{eq:NSbound0}
h \geq 0
\qquad\qquad (\hbox{Type I$_{\rm NS}$ representation with $\tilde{h} = 0$})
\end{equation}
for a unitary representation. In the non-rational regime, i.e.\ for $k\geq \frac{3}{4}$ see eq.~(\ref{kbound}), these seem to be the only requirements for unitarity; in particular, we have checked that the inner product matrix at level $\frac{1}{2}$ is positive semi-definite. 

\subsection{The Ramond sector} \label{sec:R}

The analysis in the Ramond sector is somewhat simpler since now all generators have zero modes. On the highest weight states these zero modes form an effective finite-dimensional ${\cal W}$-superalgebra\footnote{This algebra is just Zhu's algebra \cite{Zhu96}.} which we have worked out in Appendix~\ref{app:SW_kOPEs}.
It is particularly simple when expressed in the basis of $L_0$, $G_0$, $\tilde{G}_0= \frac{i}{\sqrt{15}} P_0$, $\tilde{L}_0 = - \frac{1}{5} X_0$, as well as the combinations
\begin{equation}
\tilde{K}_0=\frac{iK_0}{\sqrt{15}}+\frac{1}{4}\sqrt{\frac{1}{30k}}
\end{equation}
and
\begin{equation}
U_0 = -\frac{1}{5}\left(M_0 - \frac{G_0}{4} + i\sqrt{\frac{1}{2k}}P_0\right) \ .
\end{equation}

It will turn out that the irreducible representations of the zero mode algebra are one-dimensional or two-dimensional. In the following we shall describe the one-dimensional representations; the two-dimensional representations are explained in Appendix~\ref{app:Rrep2d}.

\subsubsection{One-dimensional representations}

For the one-dimensional representations, there are essentially two possibilities. The first possibility is 
\begin{flalign} \label{eq:Rrep380}
~&\underline{\hbox{Type I$_{\rm R}$}:} \quad h=\frac{c}{24}+\delta \ ,\qquad
\tilde{h}=\frac{3}{80}\ ,\quad \tilde{\kappa} = \sigma\tilde{\sigma}\sqrt{\frac{\delta}{30}}\ ,  &&
\end{flalign}
where $\delta$ is unconstrained, $\sigma, \tilde{\sigma}=\pm 1$, and $\tilde{\kappa}$ is the eigenvalue of $\tilde{K}_0$. For these states the fermionic generators act as 
\be
G_0=\sigma\sqrt{\delta}\ ,\qquad
\tilde{G}_0=\frac{\tilde{\sigma}}{2\sqrt{30}}\ ,\qquad U_0=0\ .
\end{equation}
The other possibility is 
\begin{flalign} \label{eq:Rrep716}
~&\underline{\hbox{Type II$_{\rm R}$}:} \qquad h=\frac{7}{16}\ ,\qquad
\tilde{h}=\frac{7}{16}\ ,\qquad \tilde{\kappa} = -\frac{7}{2}\sqrt{\frac{1}{30k}}\ , &&
\end{flalign}
where now the fermionic generators act as 
\be
G_0=\sigma \frac{1}{2}\sqrt{\frac{1}{k}}\ ,\qquad
\tilde{G}_0= - \sigma \frac{7}{2\sqrt{30}}\ ,\qquad U_0=0\ ,
\end{equation}
and $\sigma=\pm$ is again a sign.
On the face of it these results follow without imposing any null-vector relation. For example, the tricritical Ising weight $\tilde{L}_0$ is fixed by setting to zero the $\{U_0, U_0\}$ anti-commutator \eqref{eq:U0U0}, and thus there is no need for an analogue of \eqref{eq:NSfixingMMweight}. 
Actually this is because in our conventions, $N_0=0$ is automatically satisfied on the highest weight states. 

Our results are compatible with \cite[Section 5.2]{Noyvert:2002mc}, but are stronger than what was found there --- \cite{Noyvert:2002mc} worked with a different basis and did not impose all the null-vector relations. Our results should also be a finite $k$ generalisation of the Shatashvili-Vafa spectrum in \cite[Section 7]{Noyvert:2002mc}, and while this seems essentially to be true, we differ in some details.\footnote{We disagree with the third line of his equation (7.7). We almost agree with the fourth line, except that we would have put the factor of a half inside the square root, in the expression for $m$.}

As in the NS sector, the analysis up to here is independent of unitarity. Imposing unitarity requires in addition 
\begin{equation} \label{eq:RamondBound}
h\geq \frac{c}{24} = \frac{7}{16}-\frac{1}{4k} \ , 
\end{equation}
as follows from the usual $\mathcal{N}=1$ analysis. This does not impose any conditions on the representations of Type II$_{\rm R}$ (as long as $k>0$), while for representations of Type~I$_{\rm R}$ it implies that $\delta\geq 0$. In the non-rational regime ($k\geq \frac{3}{4}$) that we are primarily interested in, there may be additional constraints from unitarity, but we have not seen any evidence for this.
In particular we have checked that the inner product matrix of the level one descendants of either Type~I$_{\rm R}$ or Type~II$_{\rm R}$ primaries is positive semi-definite for $k\geq \frac{3}{4}$. This is true for any choice of the signs $\sigma,\tilde{\sigma}=\pm$ and for any $\delta\geq 0$.

\section{Implications of the symmetry algebra}\label{sec:impl}

The aim of this section is to analyse to which extent structural properties of the Shatashvili-Vafa algebra discussed in \cite{Shatashvili:1994zw}, see also \cite{deBoer:2005pt}, continue to hold for the finite $k$ generalisation $\mathcal{SW}_k(\frac{3}{2}, \frac{3}{2}, 2)$.

\subsection{From Ramond to NS primaries}\label{sec:notsospecial}

Ramond ground states saturate by definition the unitarity bound \eqref{eq:RamondBound}. They arise from our Type I$_{\rm R}$ representation \eqref{eq:Rrep380} by setting $\delta=0$. On the other hand, the Type II$_{\rm R}$ representation \eqref{eq:Rrep716} is \emph{not} a Ramond ground state unless $k\rightarrow\infty$. Nevertheless it has, even for finite $k$, the key property that makes it so central in Shatashvili-Vafa theories \cite{Shatashvili:1994zw}: \emph{it transforms trivially under $T-\tilde{T}$}. One consequence is that its fusion behaviour is tractable. In the Shatashvili-Vafa case, this has led to speculations about special NS states analogous to chiral primaries in $\mathcal{N}=2$ CFTs. In this section we perform the analogous analysis for arbitrary $k$, and identify the corresponding NS states. Unfortunately, as we shall see, they do not directly correspond to exactly marginal operators, unlike what happened in the set-up of \cite{Shatashvili:1994zw}. 

The idea of \cite{Shatashvili:1994zw} was to consider the fusion of the Ramond ground states  with the (special) representation of Type~II$_{\rm R}$, see eq.~(\ref{eq:Rrep716}), 
\be\label{fusion}
[\hbox{Type I$_{\rm R}$},h=\tfrac{c}{24}] \otimes_{\rm f} [\hbox{Type II$_{\rm R}$},h=\tfrac{7}{16}] \ . 
\ee
This can be analysed using standard fusion technology, see e.g.\ the approach of \cite{Gaberdiel:1996kf}. If we just concentrate on the tricritical Ising model, then we have the (even) fusion, see e.g.\ \cite{Eichenherr:1985cx,Sotkov:1986bv},
\be
[\tilde{h}=\tfrac{3}{80}] \otimes_{\rm f} [\tilde{h}=\tfrac{7}{16}] = [\tilde{h}=\tfrac{1}{10}] \ , 
\ee
where the two representations on the left-hand-side are in the R sector, while the fusion product is in the NS sector. Next we observe that for the ground state of the representation of Type II, we have 
\be
L_{-1} | \tfrac{7}{16} \rangle = \tilde{L}_{-1} \, | \tfrac{7}{16} \rangle \ ,
\ee
since $T-\tilde{T}$ acts trivially on this state. It then follows from \cite[eq.~(4.14)]{Gaberdiel:1996kf} that in the fusion of the above representations we have 
\be
h_3 - h_1 - h_2 = \tilde{h}_3 - \tilde{h}_1 - \tilde{h}_2 = \tfrac{1}{10} - \tfrac{3}{80} - \tfrac{7}{16}  \ , 
\ee
where $h_i$ and $\tilde{h}_i$ are the eigenvalue of $L_0$ and $\tilde{L}_0$ on the three fields in the OPE, respectively. (Here the two fields on the left-hand-side of (\ref{fusion}) have eigenvalues $(h_1,\tilde{h}_1)$ and ($h_2,\tilde{h}_2)$, while $(h_3,\tilde{h}_3)$ is the eigenvalue of the primary that appears in the fusion product.) In particular, this now allows us to compute $h_3$, and we find 
\be
h_3 = \frac{c}{24} + \frac{7}{16} + \Bigl(\frac{1}{10} - \frac{3}{80} - \frac{7}{16}  \Bigr) = \frac{1}{2} - \frac{1}{4k} \ . 
\ee
Together with the analysis of the tricritical Ising fusion, it therefore follows that 
\be\label{fusion1}
[\hbox{Type I$_{\rm R}$},h=\tfrac{c}{24}] \otimes_{\rm f} [\hbox{Type II$_{\rm R}$},h=\tfrac{7}{16}] = 
[\hbox{Type II$_{\rm NS}$},h=\tfrac{1}{2} - \tfrac{1}{4k}] \ , 
\ee
i.e.\ the fusion product is the lowest Type~II$_{\rm NS}$ NS-sector representation allowed by the unitarity bound, see eq.~(\ref{eq:NSbound110}).

Recall that the RR ground states are expected to encode the topology of the underlying manifold, and in particular one should be able to read off the Betti numbers from them. It follows from (\ref{fusion1}) that these RR ground states are in one-to-one correspondence with the special NS sector states $[\hbox{Type II$_{\rm NS}$},h=\tfrac{1}{2} - \tfrac{1}{4k}]$ (both for left- and right-movers). In the flat space limit $k\rightarrow \infty$, these states in turn give rise to exactly marginal operators (upon applying $G_{-1/2}$), but for finite $k$ this is no longer true, as is clear from dimensional analysis. Thus, the direct link between Betti numbers and the set of exactly marginal deformations is broken for finite $k$.

We mention in passing that essentially the same argument also allows us to show that 
\be\label{fusion2}
[\hbox{Type II$_{\rm R}$},h=\tfrac{7}{16}] \otimes_{\rm f} [\hbox{Type II$_{\rm R}$},h=\tfrac{7}{16}] = 
[\hbox{Type I$_{\rm NS}$},h=0] \ , 
\ee
since we have the tricritical Ising fusion 
\be
[\tilde{h}=\tfrac{7}{16}] \otimes_{\rm f} [\tilde{h}=\tfrac{7}{16}] = [\tilde{h}=0] \ .
\ee

\subsection{Chiral NS primaries and marginal deformations}\label{sec:chiralp}

As we saw in the previous section, the NS primaries $[\hbox{Type II$_{\rm NS}$},h=\tfrac{1}{2} - \tfrac{1}{4k}]$ are in one-to-one correspondence with the RR ground states. This is actually quite natural since these NS states saturate the unitarity bound of eq.~(\ref{eq:NSbound110}). 

However, as we shall explain in the following, these NS states should not be thought of as `chiral primaries'. In fact, there is another set of NS primaries that are more naturally to be identified as the chiral primaries of $\mathcal{SW}_k(\frac{3}{2}, \frac{3}{2}, 2)$, namely the $[\hbox{Type II$_{\rm NS}$},h=\tfrac{1}{2}]$ primaries with $\kappa=0$.
In particular, they are `chiral' in the sense that they are annihilated by a part of $G_{-1/2}$, see eq.~(\ref{G-null}) below and, as a consequence, they give rise to exactly marginal operators preserving the $\mathcal{SW}_k(\frac{3}{2}, \frac{3}{2}, 2)$ symmetry, see Section~\ref{sec:deformations}.

\subsubsection{Chirality} \label{sec:chirality}

Recall that in ${\cal N}=2$ theories chiral primaries in the NS sector are characterised by the property that they have an additional null-vector at level $\frac{1}{2}$. We therefore begin by analysing which NS sector representations of $\mathcal{SW}_k(\frac{3}{2}, \frac{3}{2}, 2)$ have this property. 

In general there are three $\frac{1}{2}$ descendants of any highest weight state, namely the ones created by $G_{-1/2}$, $P_{-1/2}$ and $M_{-1/2}$. As was explained in Section~\ref{sec:NS}, consistency of the algebra requires that we have the null-vector $N_{-1/2}|h,\tilde{h},\kappa\rangle=0$, see eq.~(\ref{Nexp}). For the representations of Type I$_{\rm NS}$ we have in addition $\tilde{h}=0$, and hence also the $P_{-1/2}$ descendant vanishes. (Recall that $P$ is proportional to the supercurrent of the tricritical Ising model, see eq.~(\ref{eq:TricriticalGenerators}), and $\tilde{h}=0$ means that, with respect to the tricritical Ising model, the state is in the vacuum state.) Thus for the representations of Type~I$_{\rm NS}$ there exists only one level $\frac{1}{2}$ descendant, which we may take to be 
\begin{equation} \label{eq:Gdown0}
G^\downarrow_{-1/2}\ket{h,0,0} =
\bigg(\frac{3}{2}
G_{-1/2}-\frac{i}{7}\sqrt{\frac{2}{k}}P_{-1/2}-
M_{-1/2}\bigg)\ket{h,0,0}\ ,\qquad
\tilde{L}_0 = \tfrac{1}{10}\ . 
\end{equation}
In fact, this is the unique linear combination (not equal to a null-vector) with definite $\tilde{L}_0$ eigenvalue, which turns out to be equal to $\tilde{L}_0 = \tfrac{1}{10}$. It is not difficult to check that (\ref{eq:Gdown0}) is only null provided that $h=0$. Thus the only NS-sector representation of Type~I that is `chiral' is the vacuum representation.\footnote{Alternatively, one may argue that all of these representations should be regarded as `chiral' since they all have a second null-vector because of $\tilde{h}=0$. However, then the spectrum of chiral primaries would be continuous, which is not what one usually looks for.} 

Before we continue to discuss the case of the Type~II representations, we should comment on the notation $G^\downarrow_{-1/2}$ that appears on the left-hand-side of \eqref{eq:Gdown0}. It is not difficult to see that the operator 
\begin{equation} \label{eq:Gamma}
\Gamma = G + \frac{i}{7}\sqrt{\frac{2}{k}}P 
\end{equation}
transforms in the $h_{1,2}=\frac{1}{10}$ representation with respect to the tricritical Ising model,\footnote{In the $k\rightarrow \infty$ limit $\Gamma=G$, and thus the supercurrent itself has this property. This is not true at finite $k$ though.} where we label the conformal dimensions of the tricritical Ising model via 
\be
h_{r,s} = \frac{(5r-4s)^2-1}{80}\ , \qquad 1 \leq r \leq 3 \ , \ \  1 \leq  s \leq 4 \ , \quad h_{r,s} = h_{4-r,5-s} \ ,
\ee
with $r=1,3$ for NS sector representations, and $r=2$ for R sector representations.
Thus the action by a mode of $\Gamma$ behaves as fusion with the $h_{1,2}$ representation with respect to the tricritical Ising model.  On the highest weight state of the Type I$_{\rm NS}$ representation $P_{-1/2}$ vanishes, and therefore the action of $\Gamma_{-1/2} = G_{-1/2}$ is controlled by the tricritical Ising fusion rules 
\be\label{trifus}
[h_{1,2}]\otimes_{\rm f} [h_{r,s}] = [h_{r,s-1}] \oplus [h_{r,s+1}]\ .
\ee
Following \cite{deBoer:2005pt} we can thus decompose the action of $G_{-1/2}$ into the component that moves us upwards or downwards in the Kac table, see figure~\ref{fig:Kac}, 
\be
G_{-1/2} |h,0,0\rangle = G^\uparrow_{-1/2} |h,0,0\rangle + G^\downarrow_{-1/2} |h,0,0\rangle \ . 
\ee
Since $|h,0,0\rangle$ transforms in the tricritical Ising vacuum, the resulting state must necessarily sit in the `downward' representation with $h_{1,2}=\frac{1}{10}$, i.e.\ the (potentially) non-zero state is $G^\downarrow_{-1/2} |h,0,0\rangle$, see eq.~(\ref{eq:Gdown0}). On the other hand, since there is no upward component, i.e.\ since the term $[h_{1,0}]$ in (\ref{trifus}) is absent, the corresponding state
\begin{equation}
G^\uparrow_{-1/2}\ket{h,0,0}
= \bigg(-\frac{1}{2}G_{-1/2}+\frac{i}{7}\sqrt{\frac{2}{k}}P_{-1/2}+
M_{-1/2}\bigg)\ket{h,0,0}
\end{equation}
must be null, and this is indeed the case since it can be written as a linear combination of $N_{-1/2}\ket{h,0,0}$ and $P_{-1/2} \ket{h,0,0}$. 
\smallskip

\begin{figure}[h]
\begin{center}
\begin{tikzpicture}[scale=1]
\def \q {4}

\draw[->] (0,0) -- (2,0) node[right] {$r$};
\draw[->] (0,0) -- (0,-2) node[below] {$s$};
\draw (1,-0.05) -- (1,0.05);
\draw (-0.05,-1) -- (0.05,-1);

\foreach \r in {1,...,3}{
\foreach \s in {1,...,4}{
\node [draw=black, fill=white, minimum size=1cm] at (\r,-\s) {};
}}

\foreach \r in {1,...,3}{
\foreach \s in {1,...,4}{
\pgfmathsetmacro\num{((\r * (\q+1) -\s * \q)^2 -1)/gcd(((\r * (\q+1) -\s * \q)^2 -1),4*\q*(\q+1))}
\pgfmathsetmacro\denom{4*\q*(\q+1)/gcd(((\r * (\q+1) -\s * \q)^2 -1),4*\q*(\q+1))}
\ifdim\denom pt = 1 pt\relax \node at (\r,-\s) {\large $\pgfmathprintnumber{\num}$};\fi
\ifdim\denom pt > 1 pt\relax \node at (\r,-\s) {\large $\frac{\pgfmathprintnumber{\num}}{\pgfmathprintnumber{\denom}}$};\fi
}}

\draw[->, very thick] (0.7,-1.1) -- (0.7,-1.9);
\draw[->, very thick] (0.7,-2.1) -- (0.7,-2.9);
\draw[->, very thick] (0.7,-3.1) -- (0.7,-3.9);

\draw[<-, very thick] (1.3,-1.1) -- (1.3,-1.9);
\draw[<-, very thick] (1.3,-2.1) -- (1.3,-2.9);
\draw[<-, very thick] (1.3,-3.1) -- (1.3,-3.9);

\end{tikzpicture}
\caption{Weights $h_{r,s}$ of primaries of the tricritical Ising model in its Kac table. The arrows define the projection $G^\downarrow_{-1/2}$ and $G^\uparrow_{-1/2}$ as discussed in the main text.}
\label{fig:Kac}
\end{center}
\end{figure}
\medskip

\noindent Let us now turn to the case of the Type II$_{\rm NS}$ representation, for which $\tilde{h}=h_{1,2}=\frac{1}{10}$. Again because of consistency we must have that 
\begin{align} \label{eq:Nmhalf110}
N_{-1/2}\ket{h(\kappa),\tfrac{1}{10},\kappa} =
&-\left(G_{-1/2}+2\kappa P_{-1/2}+2M_{-1/2}\right)\, \ket{h(\kappa),\tfrac{1}{10},\kappa} = 0 
\end{align}
 is null, but now we would expect in general two non-trivial descendants. In terms of $\Gamma_{-1/2}$, we therefore expect that we have an `up'-component with $\tilde{h}=h_{1,1}=0$, and a `down'-component with $\tilde{h}=h_{1,3} = \frac{3}{5}$, and we find explicitly
 \begin{align}
\Gamma^\uparrow_{-1/2}\ket{h(\kappa),\tfrac{1}{10},\kappa} & = \frac{1}{6}\big(5G_{-1/2}-2M_{-1/2}\big)\,\ket{h(\kappa),\tfrac{1}{10},\kappa}\ ,\qquad
& \tilde{L}_0 = 0\ , \nonumber \\
\Gamma^\downarrow_{-1/2}\ket{h(\kappa),\tfrac{1}{10},\kappa} & = 
\frac{1}{6}\big(G_{-1/2}+\frac{6i}{7} \, \sqrt{\frac{2}{k}} P_{-1/2} + 2M_{-1/2}\big)\,\ket{h(\kappa),\tfrac{1}{10},\kappa}\ ,
& \tilde{L}_0 = \tfrac{3}{5} \ .\nonumber 
\end{align}
Actually the $P_{-1/2}$ descendant of $\ket{h(\kappa),\tfrac{1}{10},\kappa}$ has itself eigenvalue $\tilde{L}_0 = \tfrac{3}{5}$, so using \eqref{eq:Gamma} we can decompose analogously the  supercurrent descendant $G_{-1/2}$ as
\begin{align}
G^\uparrow_{-1/2}\ket{h(\kappa),\tfrac{1}{10},\kappa}
 = 
\frac{1}{6}\big(5G_{-1/2}-2M_{-1/2}\big)&\ket{h(\kappa),\tfrac{1}{10},\kappa}\ ,\qquad
\tilde{L}_0 = 0\ , \label{eq:Gup}\\
G^\downarrow_{-1/2} \ket{h(\kappa),\tfrac{1}{10},\kappa}
=\,~
\frac{1}{6}\big(G_{-1/2}+2M_{-1/2}\big)&\ket{h(\kappa),\tfrac{1}{10},\kappa}\ ,\qquad
\tilde{L}_0 = \tfrac{3}{5}\ . \label{eq:Gplus110}
\end{align}
We can now ask for which values of $\kappa$ either of these vectors vanishes. 
As it turns out, eq.~\eqref{eq:Nmhalf110} is the only null vector at level $\tfrac{1}{2}$ in the Type II$_{\rm NS}$ representation for any $\kappa$, as long as $k\geq \tfrac{3}{4}$; so the only way a descendant can vanish is by being proportional to \eqref{eq:Nmhalf110}.\footnote{There is therefore no special shortening condition at level $\frac{1}{2}$ in the Type II$_{\rm NS}$ representation, contrary to what was claimed in \cite{deBoer:2005pt}.} It follows by direct inspection that this can only happen for the states in (\ref{eq:Gplus110}), for which   
\be\label{G-null}
G^\downarrow_{-1/2} \, \ket{h(\kappa),\tfrac{1}{10},\kappa} = 
- \frac{1}{6} N_{-1/2} \, \ket{h(\kappa),\tfrac{1}{10},\kappa}  - \frac{\kappa}{3} \, P_{-1/2} \ket{h(\kappa),\tfrac{1}{10},\kappa}  \ . 
\ee
The first term is null because of \eqref{eq:Nmhalf110}, and these states therefore become null for $\kappa=0$. We thus conclude that the `chiral primaries' of Type II$_{\rm NS}$ are 
\be\label{IIcp}
\hbox{`Chiral primaries' of Type II$_{\rm NS}$:} \qquad 
\ket{h=\tfrac{1}{2}, \tilde{h}=\tfrac{1}{10}, \kappa=0}
\ . 
\ee
Note that these states are characterised by $\kappa=0$, not just by $h=\frac{1}{2}$ --- the latter condition also has the solution $\kappa = i\sqrt{\frac{2}{k}}$, see eq.~(\ref{eq:NullConstraintNS2}).
As we shall see momentarily, see Section~\ref{sec:deformations}, the condition \eqref{IIcp} also guarantees that the $G_{-1/2}$ descendant of this chiral primary state defines an exactly marginal operator preserving the $\mathcal{SW}_k(\frac{3}{2}, \frac{3}{2}, 2)$ symmetry.

\subsubsection{Marginal deformations} \label{sec:deformations}

In this section we want to show that the $G_{-1/2}$ descendant of the chiral primary operator (\ref{IIcp}) is indeed exactly marginal. The proof follows closely \cite[p.16]{Shatashvili:1994zw}, and hinges on the fact that 
\begin{equation}\label{exmarg}
G_{-1/2}\ket{\tfrac{1}{2},\tfrac{1}{10},0} = G^\uparrow_{-1/2}\ket{\tfrac{1}{2},\tfrac{1}{10},0}
\end{equation}
transforms as the identity with respect to the tricritical Ising model ($\tilde{h} = 0$), as follows from eqs.~(\ref{G-null}) and (\ref{eq:Gup}).

More precisely Shatashvili and Vafa work with type II superstrings, for which both left- and right-movers have the symmetry $\mathcal{SW}_{k=\infty}(\frac{3}{2}, \frac{3}{2}, 2)$, and they take both the left- and right-moving state to equal (\ref{exmarg}). We shall do the same in the following, i.e.\ we shall also assume that both left- and right-moving states are (\ref{exmarg}), and then their proof applies with little change: it basically hinges on the four equations --- our notation is the same as theirs 
\begin{align}
\left(G_{-1/2} + 2M_{-1/2}\right)&\ket{\tfrac{1}{2},\tfrac{1}{10},0}=0\ ,  \label{eq1} \\
\left(M_{1/2}G_{-1/2}+2X_0\right)&\ket{\tfrac{1}{2},\tfrac{1}{10},0}=0\ , \label{eq2}\\
\left(M_{-1/2}G_{-1/2}+X_{-1}-\tfrac{1}{2}L_{-1}\right)&\ket{\tfrac{1}{2},\tfrac{1}{10},0}=0\ , \label{eq3}\\
\left(M_{-3/2}G_{-1/2}+L_{-1}X_{-1}\right)&\ket{\tfrac{1}{2},\tfrac{1}{10},0}=0\ . \label{eq4}
\end{align}
The remaining ingredients in the proof work for any $\mathcal{N}=(1,1)$ SCFT and were obtained already by Dixon \cite{Dixon:1987bg}. It therefore only remains to check that the four identities (\ref{eq1}) -- (\ref{eq4}) are also true at finite $k$. 

The first equation (\ref{eq1}) is the null-vector relation \eqref{eq:Nmhalf110}, and the third one, i.e.\ eq.~(\ref{eq3}), follows from \eqref{eq:Nmhalf110} by acting with $G_{-1/2}$ and using the algebra relations, 
\begin{equation}
\begin{split}
0&=G_{-1/2}\left(G_{-1/2} + 2M_{-1/2}\right)\ket{\tfrac{1}{2},\tfrac{1}{10},0}\\
&=\left(L_{-1} + 2(-M_{-1/2}G_{-1/2}-X_{-1})\right)\ket{\tfrac{1}{2},\tfrac{1}{10},0} \ .
\end{split}
\end{equation}
The second equation, eq.~(\ref{eq2}), also follows from a simple application of the algebra, $\{M_{1/2},G_{-1/2}\}=-2X_0$. The final equation, eq.~(\ref{eq4}), demands a bit more work. First notice that $M_{-3/2}$ anticommutes with $G_{-1/2}$, implying
\begin{equation}
M_{-3/2}G_{-1/2}+L_{-1}X_{-1}
=G_{-1/2}\left(-M_{-3/2}+G_{-1/2}X_{-1}\right) \ , 
\end{equation}
where we have also used that $G_{-1/2}^2=L_{-1}$. 
The parenthesis simplifies further thanks to $[G_{-1/2},X_{-1}]=M_{-3/2}$,
\begin{equation}
\begin{split}
\left(M_{-3/2}G_{-1/2}+L_{-1}X_{-1}\right)\ket{\tfrac{1}{2},\tfrac{1}{10},0} & = 
G_{-1/2} \left(-M_{-3/2}+G_{-1/2}X_{-1}\right)\ket{\tfrac{1}{2},\tfrac{1}{10},0} \\
& = G_{-1/2} X_{-1}G_{-1/2}\ket{\tfrac{1}{2},\tfrac{1}{10},0} \\
& = - 5 \, G_{-1/2} \tilde{L}_{-1}G^\uparrow_{-1/2}\ket{\tfrac{1}{2},\tfrac{1}{10},0} = 0 \ .
\end{split}
\end{equation}
In the penultimate step we have used that $\ket{\tfrac{1}{2},\tfrac{1}{10},0}$ is chiral, i.e.\ eq.~(\ref{exmarg}). This state is now null because $G^\uparrow_{-1/2}\ket{\tfrac{1}{10}, 0}$ transforms as a tricritical Ising vacuum, i.e.\ it has $\tilde{h}=0$, see eq.~(\ref{eq:Gup}). This completes the proof.

\subsubsection{An example of marginal deformation}

It is instructive to see how this  analysis is realised for the case of  $\text{AdS}_3\times\text{S}^3\times\mathbb{T}^4$, see Section~\ref{sec:T4}. In the $\mathbb{T}^4$ case, we have seven free fermions $\chi^1$,\ldots, $\chi^7$, all of which have $h = \frac{1}{2}$: $\chi^1$, $\chi^2$, $\chi^3$ come from $\widehat{\mathfrak{su}}(2)_k^{(1)}$, while $\chi^4$,\ldots, $\chi^7$ come from $\mathbb{T}^4$. It is easy to see that their tricritical Ising dimension equals $\tilde{h}=\frac{1}{10}$, but they actually have different $K_0$ eigenvalues, and we find that 
\begin{align}
\chi^1\,, \chi^2\,, \chi^3 \quad&\text{transform as}\quad
\ket{\tfrac{1}{2},\tfrac{1}{10},\kappa = i\sqrt{\tfrac{2}{k}}} \quad {\rm Type~II}_{\rm NS}\text{ primaries}\ ,
\\
\chi^4\,, \chi^5\,, \chi^6\,, \chi^7 \quad&\text{transform as}\quad
\ket{\tfrac{1}{2},\tfrac{1}{10},\kappa = 0}
 \quad\quad~{\rm Type~II}_{\rm NS} \text{ primaries}\ .
\end{align}
We would only expect the latter fermions to lead to exactly marginal deformations preserving $\mathcal{SW}_{k}(\frac{3}{2}, \frac{3}{2}, 2)$ --- the corresponding operators change the radii of the $\mathbb{T}^4$ --- and this fits nicely with what we saw abstractly above, namely that only the Type~II$_{\rm NS}$ primaries with $\kappa=0$ lead to exactly marginal operators. 

\subsection{A nilpotent operator?} \label{sec:nilpotent}

The idea behind the introduction of  $G_{-1/2}^\downarrow$ in \cite{deBoer:2005pt} was that it should behave as $G^+_{-1/2}$, say, for ${\cal N}=2$ theories. This is to say, it should be nilpotent, and its cohomology should give rise to the Dolbeault operator as in \cite{Lerche:1989uy}. It was also proposed as a BRST operator for a topologically twisted version of the theory, as originally suggested in \cite{Shatashvili:1994zw}.
It was shown in \cite{deBoer:2005pt} that $G_{-1/2}^\downarrow$ is indeed nilpotent in the $k\rightarrow \infty$ limit, and a conjecture was made for its cohomology. In this section we show that at least the naive definition of $G_{-1/2}^\downarrow$ is in general not nilpotent for finite $k$. Furthermore, we show in Section~\ref{sec:coho} that even in the $k\rightarrow \infty$ limit, its cohomology does not appear to capture the geometry as anticipated in \cite{deBoer:2005pt}.

Let us first analyse whether $G_{-1/2}^\downarrow$  is nilpotent at finite $k$. We recall from eq.~(\ref{eq:Gamma}) that we can write 
\begin{equation} \label{eq:DecompositionGmhalf}
G_{-1/2} = \Gamma^\downarrow_{-1/2} + \Gamma^\uparrow_{-1/2} -\frac{1}{7}\sqrt{\frac{30}{k}}\tilde{G}_{-1/2}\ , 
\end{equation}
where $\Gamma^\downarrow_{-1/2}$ and $\Gamma^\uparrow_{-1/2}$ are defined by projecting to the tricritical Ising model representations below and above, respectively, and we have expressed $P$ in terms of $\tilde{G}$, see eq.~(\ref{eq:TricriticalGenerators}). Given that $\tilde{G}$ maps the tricritical Ising representations as 
\be
\tilde{G}: \quad [h_{1,1}=0] \leftrightarrow [h_{1,4}=\tfrac{3}{2}] \ , \qquad 
[h_{1,2}=\tfrac{1}{10}] \leftrightarrow [h_{1,3}=\tfrac{3}{5}] \ , 
\ee
we can track how $G_{-1/2}$ acts between tricritical Ising conformal families using \eqref{eq:DecompositionGmhalf}. In particular we can write `up/down' projections explicitly as (writing $\alpha=-\frac{1}{7}\sqrt{\frac{30}{k}}$)
\begin{equation}  \label{eq:ExplicitDefGup}
\begin{split}
&G_{-1/2}^\uparrow\ket{[0]} = 0 \ ,
\\
&G_{-1/2}^\uparrow\ket{[\tfrac{1}{10}]} =  \mathcal{P}_{[0]}\, G_{-1/2}\ket{[\tfrac{1}{10}]} = \Gamma_{-1/2}^\uparrow\ket{[\tfrac{1}{10}]}\ ,
\\
&G_{-1/2}^\uparrow\ket{[\tfrac{3}{5}]} =  \mathcal{P}_{[\frac{1}{10}]} \, G_{-1/2}\ket{[\tfrac{3}{5}]} = \bigl(\Gamma_{-1/2}^\uparrow+\alpha\tilde{G}_{-1/2}\bigr)\ket{[\tfrac{3}{5}]}\ ,
\\
&G_{-1/2}^\uparrow\ket{[\tfrac{3}{2}]} = \mathcal{P}_{[\frac{3}{5}]}\, G_{-1/2}\ket{[\tfrac{3}{2}]} = \Gamma_{-1/2}^\uparrow\ket{[\tfrac{3}{2}]}\ , 
\end{split}
\end{equation}
and
\begin{equation}\label{eq:ExplicitDefGdown}
\begin{split} 
&G_{-1/2}^\downarrow\ket{[0]} =  \mathcal{P}_{[\frac{1}{10}]}\, G_{-1/2}\ket{[0]} = \Gamma_{-1/2}^\downarrow\ket{[0]}\ , 
\\
&G_{-1/2}^\downarrow\ket{[\tfrac{1}{10}]} =  \mathcal{P}_{[\frac{3}{5}]}\, G_{-1/2}\ket{[\tfrac{1}{10}]} = \bigl(\Gamma_{-1/2}^\downarrow+\alpha\tilde{G}_{-1/2}\bigr)\ket{[\tfrac{1}{10}]}\ ,
\\
&G_{-1/2}^\downarrow\ket{[\tfrac{3}{5}]} =  \mathcal{P}_{[\frac{3}{2}]}\, G_{-1/2}\ket{[\tfrac{3}{5}]} = \Gamma_{-1/2}^\downarrow\ket{[\tfrac{3}{5}]}\ ,
\\
&G_{-1/2}^\downarrow\ket{[\tfrac{3}{2}]} = 0\ , 
\end{split}
\end{equation}
where $\ket{[h]}$ denotes an arbitrary state in the $h_{r,s}=h$ tricritical Ising model sector, and $\mathcal{P}_{[h]}$ denotes the projection to the sector with $h_{r,s}=h$.\footnote{Here we have only considered the NS sector, i.e.\ only the cases $h_{1,1}=0$, $h_{1,2}=\frac{1}{10}$, $h_{1,3}=\frac{3}{5}$ and $h_{1,4}=\frac{3}{2}$.} However, because of the $\tilde{G}_{-1/2}$ correction term in (\ref{eq:DecompositionGmhalf}), $G_{-1/2}^\uparrow + G_{-1/2}^\downarrow \neq G_{-1/2}$, and thus the argument of \cite{deBoer:2005pt} for the nilpotency of, say, $G_{-1/2}^\downarrow$ breaks down. For example, on $\ket{[0]}$ we have
\begin{equation}
\begin{split}
(G_{-1/2})^2\ket{[0]} &=
\underbrace{\left[
\big(\Gamma_{-1/2}^\downarrow+\alpha\tilde{G}_{-1/2}\big)\Gamma_{-1/2}^\downarrow
+\alpha\Gamma_{-1/2}^\uparrow\tilde{G}_{-1/2}
\right]\ket{[0]}
}_{\in [\tfrac{3}{5}]}
\\
&\qquad\qquad +
\underbrace{\left[(\alpha\tilde{G}_{-1/2})^2 + \Gamma_{-1/2}^\uparrow\Gamma_{-1/2}^\downarrow
\right]\ket{[0]}
}_{\in [0]} \ ,
\end{split}
\end{equation}
from which we conclude
\begin{align}
L_{-1}\ket{[0]}
&= \left(\alpha^2\tilde{L}_{-1} + \Gamma_{-1/2}^\uparrow\Gamma_{-1/2}^\downarrow
\right)\ket{[0]} \ ,
\\
0
&= \left[
\left(\Gamma_{-1/2}^\downarrow+\alpha\tilde{G}_{-1/2}\right)\Gamma_{-1/2}^\downarrow
+\alpha\Gamma_{-1/2}^\uparrow\tilde{G}_{-1/2}
\right]\ket{[0]}\ , 
\end{align}
since ${L}_{-1}$ does not map one out of a given tricritical Ising representation. 
Using \eqref{eq:ExplicitDefGup} and \eqref{eq:ExplicitDefGdown} we can rewrite these as
\begin{align}
\left(\{G_{-1/2}^\uparrow, G_{-1/2}^\downarrow\}-L_{-1}\right)\ket{[0]}
&=
-\tfrac{30}{49k}\tilde{L}_{-1}\ket{[0]}\ , \label{eq:ResultsUpDownA}
\\
(G_{-1/2}^\downarrow)^2\ket{[0]}
&=
\tfrac{1}{7}\sqrt{\tfrac{30}{k}}G_{-1/2}^\uparrow\tilde{G}_{-1/2}\ket{[0]}\ .\label{eq:ResultsUpDownB}
\end{align}
Thus $G_{-1/2}^\downarrow$ is not nilpotent unless the right-hand-side of (\ref{eq:ResultsUpDownB}) vanishes. Note that (\ref{eq:ResultsUpDownB}) vanishes on the actual ground state of the $h_{1,1}=0$ tricritical Ising representation (on which $\tilde{G}_{-1/2}$ acts trivially), or in the limit $k\rightarrow \infty$. In either of these cases therefore the left-hand-side of (\ref{eq:ResultsUpDownA}) also vanishes, and we obtain the analogue of the $\mathcal{N}=2$ relation
$\{G^+_{-1/2}, G^-_{-1/2}\} = L_{-1}$. However, in general neither of these identities is true. 

For completeness, let us also mention that $(G^\uparrow_{-1/2})^2\ket{[0]}$ is trivially zero by \eqref{eq:ExplicitDefGup}, and that in the other sectors we find
\begin{align}
\left(\{G_{-1/2}^\uparrow, G_{-1/2}^\downarrow\}-L_{-1}\right)\ket{[\tfrac{1}{10}]}
&= 0\,,
\\
(G_{-1/2}^\downarrow)^2\ket{[\tfrac{1}{10}]}
&=
\tfrac{1}{7}\sqrt{\tfrac{30}{k}}\tilde{G}_{-1/2}\, G_{-1/2}^\uparrow\ket{[\tfrac{1}{10}]}\ ,
\\
\left(\{G_{-1/2}^\uparrow, G_{-1/2}^\downarrow\}-L_{-1}\right)\ket{[\tfrac{3}{5}]}
&= 0\ ,
\\
(G_{-1/2}^\uparrow)^2\ket{[\tfrac{3}{5}]}
&=
\tfrac{1}{7}\sqrt{\tfrac{30}{k}}\tilde{G}_{-1/2} \, G_{-1/2}^\downarrow\ket{[\tfrac{3}{5}]}\ ,
\\
\left(\{G_{-1/2}^\uparrow, G_{-1/2}^\downarrow\}-L_{-1}\right)\ket{[\tfrac{3}{2}]}
&= -\tfrac{30}{49k}\tilde{L}_{-1}\ket{[\tfrac{3}{2}]}\ ,
\\
(G_{-1/2}^\uparrow)^2\ket{[\tfrac{3}{2}]}
&=
\tfrac{1}{7}\sqrt{\tfrac{30}{k}}G_{-1/2}^\downarrow \, \tilde{G}_{-1/2}\ket{[\tfrac{3}{2}]}\ . \label{eq:ResultsUpDownZ}
\end{align}

Obviously, there is some arbitrariness in our definition of $G_{-1/2}^\uparrow$ and $G_{-1/2}^\downarrow$ --- while the action on highest weight states seems natural and satisfying, there are different ways in which we may extend it to descendant states. In particular, another natural definition would be to replace $G_{-1/2}^\uparrow$ of eq.~(\ref{eq:ExplicitDefGup}), by  $G_{-1/2}^\sharp$ with
\begin{equation}
\begin{split} \label{eq:ExplicitDefGsharp}
&G_{-1/2}^\sharp\ket{[0]} = \mathcal{P}_{[\frac{3}{2}]}G_{-1/2}\ket{[0]}=\alpha\tilde{G}_{-1/2}\, \ket{[0]} \ , 
\\
&G_{-1/2}^\sharp\ket{[\tfrac{1}{10}]} =  \mathcal{P}_{[0]}G_{-1/2}\ket{[\tfrac{1}{10}]} = \Gamma_{-1/2}^\uparrow\ket{[\tfrac{1}{10}]}\ ,
\\
&G_{-1/2}^\sharp\ket{[\tfrac{3}{5}]} =  \mathcal{P}_{[\frac{1}{10}]}G_{-1/2}\ket{[\tfrac{3}{5}]} = \Bigl(\Gamma_{-1/2}^\uparrow +\alpha\tilde{G}_{-1/2}\Bigr)\ket{[\tfrac{3}{5}]}\ ,
\\
&G_{-1/2}^\sharp\ket{[\tfrac{3}{2}]} = \mathcal{P}_{[\frac{3}{5}]}G_{-1/2}\ket{[\tfrac{3}{2}]} = \Gamma_{-1/2}^\uparrow\ket{[\tfrac{3}{2}]}\ , 
\end{split}
\end{equation}
i.e.\ to modify the action on $\ket{[0]}$. Similarly, for $G_{-1/2}^\downarrow$, cf.\ eq.~(\ref{eq:ExplicitDefGdown}), we modify the action on $\ket{[\tfrac{3}{2}]}$ by defining $G_{-1/2}^\flat$ via 

\begin{equation}
\begin{split} \label{eq:ExplicitDefGflat}
&G_{-1/2}^\flat\ket{[0]} =  \mathcal{P}_{[\frac{1}{10}]}G_{-1/2}\ket{[0]} = \Gamma_{-1/2}^\downarrow\ket{[0]}\ ,
\\
&G_{-1/2}^\flat\ket{[\tfrac{1}{10}]} =  \mathcal{P}_{[\frac{3}{5}]}G_{-1/2}\ket{[\tfrac{1}{10}]} = \Bigl(\Gamma_{-1/2}^\downarrow+\alpha\tilde{G}_{-1/2}\Bigr)\ket{[\tfrac{1}{10}]}\ ,
\\
&G_{-1/2}^\flat\ket{[\tfrac{3}{5}]} =  \mathcal{P}_{[\frac{3}{2}]}G_{-1/2}\ket{[\tfrac{3}{5}]} = \Gamma_{-1/2}^\downarrow\ket{[\tfrac{3}{5}]}\ ,
\\
&G_{-1/2}^\flat\ket{[\tfrac{3}{2}]} = \mathcal{P}_{[0]}G_{-1/2}\ket{[\tfrac{3}{2}]}=\alpha\tilde{G}_{-1/2}\ket{[\tfrac{3}{2}]}\ .
\end{split}
\end{equation}
This agrees with the original definition on $\mathcal{SW}_k(\tfrac{3}{2}, \tfrac{3}{2}, 2)$ 
highest weight states since $\tilde{G}_{-1/2}=0$ on Type I$_{\rm NS}$ primaries and none of the 
$\mathcal{SW}_k(\tfrac{3}{2}, \tfrac{3}{2}, 2)$ primary states is in the sector $[\tfrac{3}{2}]$, as follows from our classification of highest weight representations in Section~\ref{sec:NS}. 
The new definition has the advantage that 
\begin{equation}
G_{-1/2} = G^\flat_{-1/2} + G^\sharp_{-1/2}\ , 
\end{equation}
and that the analogue of \eqref{eq:ResultsUpDownA} -- \eqref{eq:ResultsUpDownZ} is now simply
\begin{equation} \label{eq:FlatSharpRelations}
(G^\flat_{-1/2})^2 + (G^\sharp_{-1/2})^2 = 0 \ , \qquad\qquad
\{G^\flat_{-1/2}, G^\sharp_{-1/2}\}=L_{-1}\ .
\end{equation}
However, the individual operators are still not nilpotent; for example, on the state $P=P_{-3/2} |0\rangle$ we find 
\begin{equation}
(G^\flat_{-1/2})^2 P
= \tfrac{30i}{7}\sqrt{\tfrac{2}{k}}\Bigl(-\tfrac{1}{5}M+\tfrac{i}{14}\sqrt{\tfrac{2}{k}}\, \partial P\Bigr)
\neq 0\ .
\end{equation}

\subsection{Cohomology in flat space limit}\label{sec:coho}

While for finite $k$ we have not managed to find a natural nilpotent operator, generalising the construction of $G^\downarrow_{-1/2}$ from \cite{deBoer:2005pt}, in the large $k$ limit we have 
\be
G^\downarrow_{-1/2} \stackrel{k\rightarrow \infty}{=} G^\flat_{-1/2} \ , \qquad (G^\downarrow_{-1/2})^2 \stackrel{k\rightarrow \infty}{=} 0 \ .
\ee
Furthermore, either of these operators agrees with the proposal of \cite{deBoer:2005pt} in this limit. 

It is then a natural idea that the cohomology of this operator should capture the topology of the underlying background, in particular given the link between RR ground states and $G^\downarrow_{-1/2}$-chiral primaries for $k\rightarrow\infty$, see Section~\ref{sec:notsospecial} above. This was suggested in \cite{deBoer:2005pt}, and, in particular, they made the conjecture that the non-trivial representatives of the $G^\downarrow_{-1/2}$ cohomology  can be described by highest weight states, see the end of Section~4.4.1 of \cite{deBoer:2005pt}. We have studied this question for the simple example corresponding to flat space, i.e.\ using the embedding of 
\begin{equation}
\mathcal{SW}_{k=\infty}(\tfrac{3}{2}, \tfrac{3}{2}, 2)
~\subset~
\widehat{\mathfrak{u}}(1)^{ 7}\oplus \text{Fer}^7\ .
\end{equation}
In this example, the explicit expressions for $P$ and $G$ are, cf.\ eq.~(\ref{GT4}) in Appendix~\ref{app:Dn},
\begin{align}
G &=  \sum_{i=1}^7 {\mathcal{K}^i\chi^i}\ ,\\
P &= {\chi^1\chi^2\chi^3}+\chi^1({\chi^4\chi^5}+{\chi^6\chi^7})+\chi^2({\chi^4\chi^6}-{\chi^5\chi^7})-\chi^3({\chi^4\chi^7}+{\chi^5\chi^6})\ , \nonumber
\end{align}
where we have rescaled the fields so that they satisfy, cf.\ eq.~(\ref{eq:T4fields}) 
\be
\begin{aligned}
&\mathcal{K}^i\text{ (bosonic)} & \qquad \mathcal{K}^i(z)\mathcal{K}^j(w) \sim \frac{1 }{(z-w)^2} \, \delta^{ij}
\\
&\chi^i\text{ (fermionic)} & \qquad \chi^i(z)\chi^j(w) \sim \frac{1 }{(z-w)^2} \, \delta^{ij}
\end{aligned}
\qquad\quad \text{for }i, j\in\{1,2,\ldots,7\}\ .
\ee

We have studied the cohomology of $G^\downarrow_{-1/2}$ in this example, and we have found that the state $G=G_{-3/2} |0\rangle$ gives rise to a non-trivial cohomology class. Indeed, it is easy to see that it is closed,
\begin{equation}
G_{-1/2}^\downarrow G
= \mathcal{P}_{[\frac{3}{5}]} G_{-1/2}G
= \mathcal{P}_{[\frac{3}{5}]} (2T)
= 0\ , 
\end{equation}
where we have used that $G$ defines a tricritical Ising primary with $h_{1,2}=\frac{1}{10}$, see also eq.~(\ref{eq:Gamma}). 
To show that $G$ is not exact, we note that it is a singlet under the natural action of G$_2$, which therefore also commutes with $G^\downarrow_{-1/2}$. Thus we need to look for a G$_2$-singlet of weight $h = 1$ and $\tilde{h}=0$ on which $G^\downarrow_{-1/2}$ yields $G$. But one shows by explicit inspection that no such G$_2$ singlet exists: the 28 states with $h = 1$ are spanned by the bosonic currents $\mathcal{K}^i$ and the fermion bilinears $\chi^i\chi^j$. The former transform in the $\mathbf{7}$-dimensional irreducible representation of G$_2$, while the latter sit in the $\mathbf{7}\oplus \mathbf{14}$. This proves that $G$ defines a non-trivial element of the $G^\downarrow_{-1/2}$-cohomology. On the other hand, it is evidently not primary with respect to  $\mathcal{SW}_{k=\infty}(\tfrac{3}{2}, \tfrac{3}{2}, 2)$.

We have similarly found another non-trivial cohomology class. There are $4222$ independent $(\widehat{\mathfrak{u}}(1)^{ 7}\oplus \text{Fer}^7)$-descendants of the vacuum with $h = \frac{7}{2}$. With respect to the tricritical Ising algebra, they split into a $4024$-dimensional space in $[\frac{1}{10}]$, and a $198$-dimensional space in $[\frac{3}{2}]$. The latter are trivially $G_{-1/2}^\downarrow$-closed by \eqref{eq:ExplicitDefGdown}, and they transform with respect to the G$_2$ symmetry as 
\begin{equation}\label{198}
198 = 4\cdot \bm{1} \oplus 3\cdot \bm{7} \oplus 2 \cdot \bm{14}  \oplus 3 \cdot \bm{27} \oplus \bm{64}
\qquad (h = \tfrac{7}{2},~\tilde{h}\in \tfrac{3}{2}+\mathbb{N}) \ .
\end{equation}
We can similarly decompose the $699$-dimensional space of $h=3$ states on which $G_{-1/2}^\downarrow$ could possibly produce these $198$ states as
\begin{equation}\label{699}
699 = 3 \cdot\bm{1} \oplus 11\cdot \bm{7} \oplus 6 \cdot \bm{14} \oplus 7 \cdot \bm{27} \oplus 3\cdot \bm{64} \oplus 2\cdot \bm{77} \qquad (h = 3,~\tilde{h}\in \tfrac{3}{5}+\mathbb{N}) \ .
\end{equation}
Since the $198$ states contain $4$ singlets with respect to G$_2$, but the $699$ states in (\ref{699}) only contain $3$ singlets, it is clear that at least one singlet state among the $198$ states in (\ref{198}) is not $G_{-1/2}^\downarrow$-exact. This is another example of a non-trivial cohomology class that is not primary with respect to  $\mathcal{SW}_{k=\infty}(\tfrac{3}{2}, \tfrac{3}{2}, 2)$.\footnote{A full analysis of the cohomology is complicated in view of the large number of states. With this somewhat coarse method the above states are the only non-trivial cohomology classes that we could identify up to $h = \frac{7}{2}$ (excluding the elements already identified in \cite{deBoer:2005pt}).}

These results demonstrate that the cohomology of $G^\downarrow_{-1/2}$ does not just consist of $\mathcal{SW}_{k=\infty}(\tfrac{3}{2}, \tfrac{3}{2}, 2)$ primaries, and furthermore, that it does not seem to be directly related to the underlying geometry of the background. Obviously, these statements apply to this specific definition of $G^\downarrow_{-1/2}$, and they do not preclude that some modified definition may work better. It may be interesting to explore this further. 
\medskip

In this context it may also be worth pointing out that \cite{Shatashvili:1994zw} gave another argument in favour of the topological nature of this theory which was based on a Coulomb gas formalism, see Section~5 of that paper. While these arguments may not be completely watertight --- see in particular \cite[Section~4.2]{deBoer:2005pt}  --- the argument displays intriguing cancellations: it predicts a vanishing central charge after the twist as well as a vanishing dimension for all the chiral primary states, i.e.\ for the primary states that are non-trivial in the $G^\downarrow_{-1/2}$ cohomology. It was recently shown in \cite{Fiset:2020lmg} that these curious cancellations actually occur in any unitary $\mathcal{SW}(\frac{3}{2}, 2)$ CFT. It would be interesting to clarify the physical context in which the Coulomb gas argument is legitimate, in order to explain these cancellations.

In any case, repeating verbatim the derivation in \cite[Section~5]{Shatashvili:1994zw} to our deformed algebra $\mathcal{SW}_k(\frac{3}{2}, \frac{3}{2}, 2)$, the twisted central charge reads
\begin{equation}
c_{\text{twisted}} = -\frac{98}{10} + \left(\frac{98}{10} - \frac{6}{k}\right) = - \frac{6}{k}\ ,
\end{equation}
thus suggesting that the finite $k$ theory may not allow for a topological twist. Again, this does not preclude that one may be able to define an interesting topologically twisted theory by some other method, but it does seem to confirm our findings from above. 

\section{Discussion}\label{sec:discussion}

In this paper we have described the natural deformation $\mathcal{SW}_k(\tfrac{3}{2}, \tfrac{3}{2}, 2)$ of the Shatash\-vi\-li-Vafa chiral algebra for superstrings in backgrounds of the form $\text{AdS}_3\times \mathcal{M}_7$. The deformation parameter $\sqrt{k}>0$ describes the $\text{AdS}_3$ radius measured in string units, and $k\rightarrow\infty$ reproduces the Shatashvili-Vafa result where $\text{AdS}_3$ is replaced by 3-dimensional Minkowski space $\mathbb{M}_3$. The parameter $k$ also controls the central charge of the overall $\mathcal{N}=1$ algebra as $c=\frac{21}{2}-\frac{6}{k}$. We have shown that this algebra appears in the world-sheet description of (type II) strings for the backgrounds,
\be\label{backg}
{\rm AdS}_3 \times {\rm S}^3 \times \mathbb{T}^4 \ , \qquad 
{\rm AdS}_3 \times {\rm S}^3 \times \text{K3} \ , \qquad 
{\rm AdS}_3 \times \bigl( {\rm S}^3 \times \mathbb{T}^4 \bigr) / D_n \ , 
\ee
where we have assumed that the ${\rm AdS}_3$ flux is purely NS-NS. 

We then examined general aspects of this algebra. While there are rational coset models for some discrete choices $\frac{2}{3}\leq k < \frac{3}{4}$, see eqs.~\eqref{coset} and \eqref{eq:cNoyvert} with $k_1=1$, we mainly focused on aspects that are valid in the non-rational regime $k\geq \frac{3}{4}$. In particular, we have shown that the algebra always contains the tricritical Ising minimal model subalgebra \eqref{eq:TricriticalGenerators}, and that it requires the presence of a null-vector \eqref{eq:N} for consistency. This null vector also allowed to us to describe the spectrum of Neveu-Schwarz and Ramond highest weight states systematically and without direct reference to unitarity. 

We finally studied whether the apparent similarities between the Shatashvili-Vafa algebra and $\mathcal{N}=(2,2)$ conformal field theories, discussed in \cite{Shatashvili:1994zw} and \cite{deBoer:2005pt}, survive the deformation to finite $k$. This is the case to a degree: some NS primaries, see especially eq.~\eqref{IIcp}, can indeed be considered `chiral primaries' of $\mathcal{SW}_k(\tfrac{3}{2}, \tfrac{3}{2}, 2)$ in that they are annihilated by a projection $G_{-1/2}^\downarrow$ of the supercharge to an appropriate tricritical Ising representation (much like $\mathcal{N}=2$ chiral primaries are annihilated by a projection $G^+_{-1/2}$ to a definite $\widehat{\mathfrak{u}}(1)$ representation). As a consequence they describe the exactly marginal perturbations of the theory, see Section~\ref{sec:deformations}. However some features are lost at finite $k$, such as the correspondence with Betti numbers of the target space via RR ground states, see Section~\ref{sec:notsospecial}. There is also no more obvious extension \cite{deBoer:2005pt} of $G_{-1/2}^\downarrow$ to descendants satisfying $(G_{-1/2}^\downarrow)^2 = 0$, although we do find a way to achieve some weaker relations, eq.~\eqref{eq:FlatSharpRelations}, reminiscent of $\mathcal{N}=2$ anti-commutators. We presented evidence however that, even in the Shatashvili-Vafa limit $k\rightarrow\infty$, the naive nilpotent  $G_{-1/2}^\downarrow$ does not live up to the expectations in \cite{deBoer:2005pt}. In particular, in the background $\mathbb{M}_3\times\mathbb{T}^7$, its cohomology is not restricted to $\mathcal{SW}_k(\tfrac{3}{2}, \tfrac{3}{2}, 2)$ chiral primaries, and it does not seem to carry any useful geometric interpretation. It may be interesting to see whether one can modify the definition of $G_{-1/2}^\downarrow$ so that its cohomology has a more direct geometric interpretation. 
\medskip

It would be interesting to construct the spacetime supersymmetry generator from the worldsheet data $\widehat{\mathfrak{su}}(2)_k^{(1)}\oplus\mathcal{SW}_k(\frac{3}{2}, \frac{3}{2}, 2)$, in the spirit of \cite{Banks:1987cy}. In the flat space limit $k\rightarrow\infty$ an ansatz was proposed in \cite{Shatashvili:1994zw} for the supersymmetry current in $\mathbb{M}_3$, and it seems a priori also valid at finite $k$. It is given by the product of a spin field from the superconformal ghost sector, a spin field from the three free fermions --- which are present both in $\mathbb{M}_3$ and in the WZW description of $\text{AdS}_3$ --- and finally a field $\Sigma$ from $\mathcal{SW}_k(\frac{3}{2}, \frac{3}{2}, 2)$, which must have weight $h=\frac{7}{16}$ in order for the total dimension to be $1$,
\begin{equation}
1 = \underbrace{\left(\frac{3}{8}\right)}_{\text{Ghosts}} + \underbrace{3\cdot\left(\frac{1}{16}\right)}_{\text{Fer}^3} + \underbrace{\left(\frac{7}{16}\right)}_{\mathcal{SW}_k(\frac{3}{2}, \frac{3}{2}, 2)} \ .
\end{equation}
The candidate for $\Sigma$ suggested in \cite{Shatashvili:1994zw} is the Type II$_\text{R}$ highest weight state, whose fusion rules, see Section~\ref{sec:notsospecial}, were argued to suggest a version of `spectral flow' between the Ramond and Neveu--Schwarz sectors.\footnote{One could also consider using Type I$_\text{R}$ highest weight states with $\delta = \frac{1}{4k}$, but fusion rules are then not tractable.} 
It would be interesting to work out this idea in more detail. In particular, one should also define a consistent fermion number and GSO projection, which are particularly subtle given the absence of chiral spinors in odd-dimensional spaces, see \cite{deBoer:2005pt}. The curved nature of AdS$_3$ also makes this calculation somewhat unusual.

It would also be interesting to consider other string backgrounds. 
While the $\mathcal{SW}_k(\tfrac{3}{2}, \tfrac{3}{2}, 2)$ algebra appears in the world-sheet theory of the backgrounds of eq.~(\ref{backg}), we did not find an obvious ansatz analogous to $P$ in \eqref{eq:PforK3} that would generate $\mathcal{SW}_k(\tfrac{3}{2}, \tfrac{3}{2}, 2)$ in the world-sheet WZW description of $\text{AdS}_3\times \text{S}^3 \times \text{S}^3 \times \text{S}^1$ \cite{Elitzur:1998mm}, although we have not ruled this out either.\footnote{In view of eq.~(\ref{k=3/4}) one possible explanation could be that the relevant algebra in that case is a more general algebra of the form $\mathcal{SW}(\tfrac{3}{2}, \tfrac{3}{2}, 2)$.}
 More generally, one may analyse the backgrounds of the form  \cite{Giveon:1999zm, Giveon:2005mi}  
\be\label{Giveon}
{\rm AdS}_3 \times {\rm S}^1 \times \mathsf{M}_n  \ , 
\ee
where $\mathsf{M}_n$ is described by the $n$'th ${\cal N}=2$ minimal model (with $n\geq 3$), and the level $k$ of the ${\rm AdS}_3$ factor (in the WZW description) is of the form $k=\frac{n}{n+1}$, see \cite[eq.~(2.11)]{Giveon:2005mi}. (The case $n=2$ corresponds to the situation where the central charge of the minimal model describing $\mathsf{M}_n$ vanishes.) While 
$\mathcal{SW}_k(\frac{3}{2},\frac{3}{2}, 2)$ is generically not a subalgebra of this world-sheet chiral algebra --- the only exceptional case arises for the $n=3$ ${\cal N}=2$ minimal model, see below --- the world-sheet theory always contains a more general algebra of the form ${\cal SW}(\tfrac{3}{2},\tfrac{3}{2},2)$, see eq.~(\ref{coset}), with $k_1=n-2$, 
\begin{equation}\label{k=3/4}
\left. \mathcal{SW}(\tfrac{3}{2},\tfrac{3}{2}, 2) \right|_{k_1=n-2}
\subset (\mathcal{N}=2)_{n} \oplus \widehat{\mathfrak{u}}(1) \oplus \text{Fer} \ .
\end{equation}
Here the total ${\cal N}=1$ supercharge is simply $G=G^{(\mathcal{N}=2)} + j\psi$, where $j$ and $\psi$ are the free boson and fermion coming from the ${\rm S}^1$, respectively. Furthermore, the other (primitive) generator $P$ of $\mathcal{SW}(\tfrac{3}{2},\tfrac{3}{2}, 2)$ is also invariant under the $\mathbb{Z}_2$ automorphism of \cite{Giveon:1999jg,Berenstein:1999gj}, under which the backgrounds of eq.~(\ref{Giveon}) have ${\cal N}=1$ spacetime  supersymmetry. Thus the construction of eq.~(\ref{k=3/4}) realises in some sense the expectations of \cite{Giveon:1999jg} that an analogue of the Shatashvili-Vafa algebra should exist for these backgrounds. However, unless $n=3$, the resulting algebra does not have the tricritical Ising model as a subalgebra, i.e.\ it does not agree with ${\cal SW}_k(\tfrac{3}{2},\tfrac{3}{2},2)$. 
It would be interesting to understand the physical implications of this fact. 

We should also note that the limit $n\rightarrow \infty$ leads to $k\rightarrow 1$, which is the special level where the string theory is exactly dual to a $2$d symmetric orbifold \cite{Gaberdiel:2018rqv,Eberhardt:2018ouy,Eberhardt:2019ywk}, see also \cite{Giveon:2005mi}, and it may be interesting to explore that. More generally, it may be interesting to see whether the constraints from  $\mathcal{SW}_k(\frac{3}{2},\frac{3}{2}, 2)$ (or possibly $\mathcal{SW}(\frac{3}{2},\frac{3}{2}, 2)$ in the more general case of eq.~(\ref{Giveon}))
may lead to any interesting insights into the structure of the dual CFT.

\section*{Acknowledgements}

The work of MAF is supported by an SNF grant, and we are both supported by the NCCR SwissMAP that is also funded by the Swiss National Science Foundation. MAF thanks  
Sebastjan Cizel, Mateo Galdeano Solans, Stefano Massai, and 
Xenia de la Ossa for discussions. 

\appendix

\section[Conventions for $\widehat{\mathfrak{su}}(2)^{(1)}_k$]{Conventions for $\boldsymbol{\widehat{\mathfrak{su}}(2)^{(1)}_k}$} \label{sec:su(2)conventions}

In this appendix we specify our conventions for $\widehat{\mathfrak{su}}(2)^{(1)}_k$. The generators before decoupling the fermions are denoted by 
$K^i$ (bosonic), and $\chi^j$ (fermionic), where $i,j\in \{1,2,3\}$. The defining relations are 
\begin{align}
K^i(z) K^j(w) \sim \frac{k\, \delta^{ij}}{2 (z-w)^2} &+ \frac{i\epsilon^{ij}{}_m K^m(w)}{(z-w)}\ ,
\qquad\qquad
\chi^i(z) \chi^j(w) \sim \frac{k\delta^{ij}}{2 (z-w)}\ ,\\
&\quad
K^i(z) \chi^j(w) \sim \frac{i\epsilon^{ij}{}_m \chi^m(w)}{(z-w)}\ .
\end{align}

\noindent For $k\neq 0$, we can decouple the fermions by introducing the generators 
\begin{equation}
\mathcal{K}^i = K^i + \frac{2i}{k}\epsilon^i{}_{jk}\chi^j\chi^k
\ ,
\end{equation} 
which satisfy
\begin{equation}
\mathcal{K}^i(z) \mathcal{K}^j(w) \sim \frac{(k-2)\delta^{ij}}{2 (z-w)^2} + \frac{i\epsilon^{ij}{}_m\mathcal{K}^m(w)}{(z-w)}\ ,
\qquad\qquad
\mathcal{K}^i(z) \chi^j(w) \sim 0\ .
\end{equation}

\subsection*{Superconformal symmetry}

The ${\cal N}=1$ superaffine algebra contains an $\mathcal{N}=1$ superconformal algebra with generators 
\begin{equation} \label{eq:TGS3}
T^{\text{S}^3}
=\sum_{i=1}^3\left(\frac{1}{k}\mathcal{K}^i\mathcal{K}^i+\frac{1}{k}\partial\chi^i\chi^i\right)\,,
\qquad
G^{\text{S}^3}
=\sum_{i=1}^3\frac{2}{k}\mathcal{K}^i\chi^i -\frac{4i}{k^2}\chi^1\chi^2\chi^3\,.
\end{equation}
The central charge of this $\mathcal{N}=1$ superconformal algebra is 
\begin{equation}
c=\frac{9}{2}-\frac{6}{k}\ .
\end{equation}

\section{The algebra $\mathcal{SW}_{k}(\frac{3}{2}, \frac{3}{2}, 2)$} \label{app:SW_kOPEs}

\subsection{OPEs}

Generators: $T$, $X$, $K=G_{-1/2}P$ (bosonic) and $G$, $P$, $M=G_{-1/2}X$ (fermionic).

\begin{equation}
c=\frac{21}{2}-\frac{6}{k}
\end{equation}

\begin{align}
T(z)T(0) &\sim
\frac{c}{2z^4}
+\frac{2T}{z^2}
+\frac{\partial T}{z}
\\
T(z)G(0) &\sim
\frac{3G}{2z^2}
+\frac{\partial G}{z}
\\
G(z)G(0) &\sim
\frac{2c}{3z^3}
+\frac{2T}{z}
\\
T(z)P(0) &\sim
\frac{3P}{2z^2}
+\frac{\partial P}{z}
\\
T(z)X(0) &\sim
-\frac{7}{4z^4}
+\frac{2X}{z^2}
+\frac{\partial X}{z}
\\
G(z)P(0) &\sim
\sqrt{\frac{2}{k}}\frac{i}{z^3}
+\frac{K}{z}
\\
G(z)X(0) &\sim
\frac{1}{z^2}\Bigl(-\tfrac{1}{2}G+i\sqrt{\tfrac{2}{k}}P\Bigr)
+\frac{M}{z}
\end{align}

\begin{align}
P(z)P(0) &\sim
-\frac{7}{z^3}
+\frac{6X}{z}
\\
P(z)X(0) &\sim
-\frac{15 P}{2 z^2}
-\frac{5 \partial P}{2z}
\\
X(z)X(0) &\sim
\frac{35}{4z^4}
-\frac{10X}{z^2}
-\frac{5\partial X}{z}
\end{align}

\begin{align}
T(z)K(0) &\sim
\frac{3i}{z^4}\sqrt{\frac{1}{2k}}
+\frac{2K}{z^2}
+\frac{\partial K}{z}
\\
T(z)M(0) &\sim
\frac{1}{z^3}\Bigl(-\tfrac{1}{2}G+i\sqrt{\tfrac{2}{k}}P\Bigr)
+\frac{5M}{z^2}
+\frac{\partial M}{z}
\\
G(z)K(0) &\sim
\frac{3P}{z^2}
+\frac{\partial P}{z}
\\
G(z)M(0) &\sim
-\frac{7}{2z^4}
+\frac{1}{z^2}\Bigl(4X+T-i\sqrt{\tfrac{2}{k}}K\Bigr)
+\frac{\partial X}{z}
\end{align}

\begin{align}
P(z)K(0) &\sim
\frac{1}{z^2}\Bigl(-3G+6i\sqrt{\tfrac{2}{k}}P\Bigr)
+\frac{3}{z}\Bigl(-M-\tfrac{1}{2}\partial G +i\sqrt{\tfrac{2}{k}}\partial P\Bigr) 
\\
P(z)M(0) &\sim
-\frac{15i}{z^4}\sqrt{\frac{1}{2k}}
+\frac{1}{z^2}\Bigl(\tfrac{9}{2}K+6i\sqrt{\tfrac{2}{k}}X\Bigr)
+\frac{1}{z}\Bigl(3PG-\tfrac{1}{2}\partial K +3i\sqrt{\tfrac{2}{k}}\partial X\Bigr) \nonumber
\\
X(z)K(0) &\sim
-\frac{15i}{z^4}\sqrt{\frac{1}{2k}}
+\frac{1}{z^2}\Bigl(-3K+6i\sqrt{\tfrac{2}{k}}X\Bigr)
+\frac{1}{z}\Bigl(-3PG+3i\sqrt{\tfrac{2}{k}}\partial X\Bigr)
\\
X(z)M(0) &\sim
\frac{1}{z^3}\Bigl(-\tfrac{9}{2}G-6i\sqrt{\tfrac{2}{k}}P\Bigr)
+\frac{1}{z^2}\Bigl(-5M-\tfrac{9}{4}\partial G-3i\sqrt{\tfrac{2}{k}}\partial P\Bigr)\nonumber\\
&\qquad+\frac{1}{z}\Bigl(4XG+\tfrac{1}{2}\partial M+\tfrac{1}{4}\partial^2 G-\tfrac{27i}{7}\sqrt{\tfrac{1}{2k}}\partial^2 P+\tfrac{4i}{7}\sqrt{\tfrac{2}{k}}XP\Bigr)
\\
K(z)K(0) &\sim
-\frac{21}{z^4}
+\frac{6}{z^2}\Bigl(-T+X+i\sqrt{\tfrac{2}{k}}K\Bigr)
+\frac{3}{z^2}\Bigl(-\partial T+\partial X+i\sqrt{\tfrac{2}{k}}\partial K\Bigr)
\\[4pt]
K(z)M(0) &\sim
-\frac{15 P}{z^3}
+\frac{1}{z^2}\Bigl(-\tfrac{11}{2}\partial P +6i\sqrt{\tfrac{2}{k}}M\Bigr)
+\frac{3}{z}\Bigl(GK-2TP+i\sqrt{\tfrac{2}{k}}\partial M\Bigr) \nonumber
\\
M(z)M(0) &\sim
-\frac{35}{z^5}
+\frac{1}{z^3}\bigl(-9T+20X-6i\sqrt{\tfrac{2}{k}}K\bigr)
+\frac{1}{z^2}\bigl(-\tfrac{9}{2}\partial T+10\partial X-3i\sqrt{\tfrac{2}{k}}\partial K\bigr) \nonumber \\
&\qquad + \frac{1}{z}\Bigl(\tfrac{11}{5}\partial^2 X -\tfrac{5}{2}\partial^2 T -6GM+12 TX+\tfrac{2}{5}XX-KK\Bigr) \label{eq:MM}
\end{align}

To make contact with the literature on the Shatashvili-Vafa algebra, it is useful to know that the order 1 pole in our $M(z)M(0)$ OPE can also be given as follows by exploiting null fields at level $4$:
\begin{equation}
\begin{split}
&-4GM+8TX+\frac{3}{2} \partial^2 (X -T) \\ 
&\qquad -\frac{i}{13} \sqrt{\frac{2}{k}} \left(21   \partial GP+18  G\partial P-6i \sqrt{\frac{2}{k}} \partial PP-4 XK+10 PM\right)\,.
\end{split}
\end{equation}

\subsection{Mode algebra}

\begin{align}
[L_m, L_n] &=
(m-n)L_{m+n}+\frac{c}{12}m(m^2-1)\delta_{m+n,0}
\\
[L_n, G_r] &=
\bigl(\tfrac{n}{2}-r\bigr)G_{n+r}
\\
\{G_r, G_s\} &=
2L_{r+s}+\frac{c}{3}\bigl(r^2-\tfrac{1}{4}\bigr)\delta_{r+s,0}
\\
[L_n, P_r] &= 
\bigl(\tfrac{n}{2}-r\bigr)P_{n+r}
\\
[L_m, X_n] &=
(m-n)X_{m+n}-\frac{7}{24}m(m^2-1)\delta_{m+n,0}
\\
\{G_r, P_s\} &=
K_{r+s}+i\sqrt{\frac{1}{2k}}\bigl(r^2-\tfrac{1}{4}\bigr)\delta_{r+s,0}
\\
[G_r, X_n] &=
M_{r+n}-\bigl(r+\tfrac{1}{2}\bigr) \Bigl(\tfrac{1}{2}G_{r+n}-i\sqrt{\tfrac{2}{k}}P_{r+n}\Bigr)
\\
\{P_r, P_s\} &=
6X_{r+s}-\frac{7}{2}\big(r^2-\tfrac{1}{4}\big)\delta_{r+s,0}
\\
[P_r, X_n] &=
-5\big(r-\tfrac{n}{2}\big)P_{r+n}
\\
[X_m, X_n] &=
-5(m-n)X_{m+n}+\frac{35}{24}m(m^2-1)\delta_{m+n,0}
\end{align}
\begin{align}
[L_m, K_n] &=
(m-n)K_{m+n}+\frac{i}{2}\sqrt{\frac{1}{2k}}m(m^2-1)\delta_{m+n,0}
\\
[L_n, M_r] &=
\big(\tfrac{3n}{2}-r\big)M_{n+r}-\frac{1}{4}n(n+1)G_{n+r}+i\sqrt{\frac{1}{2k}}n(n+1)P_{n+r}
\\
[G_r, K_n] &=
(2r-n)P_{r+n}
\\
\{G_r, M_s\} &=
\big(r+\tfrac{1}{2}\big)L_{r+s}
+(3r-s)X_{r+s}
-i\sqrt{\frac{2}{k}}\big(r+\tfrac{1}{2}\big)K_{r+s} \nonumber\\
&\qquad-\frac{7}{12}\big(r^2-\tfrac{1}{4}\big)\big(r-\tfrac{3}{2}\big)\delta_{r+s,0}
\\
[P_r, K_n] &=
-3M_{r+n}-\frac{3}{2}\big(r-n-\tfrac{1}{2}\big)G_{r+n}+3i\sqrt{\frac{2}{k}}\big(r-n-\tfrac{1}{2}\big)P_{r+n}
\\
\{P_r, M_s\} &=
\big(2r-\tfrac{5s}{2}-\tfrac{11}{4}\big)K_{r+s}
+3i(r-s-1)\sqrt{\frac{2}{k}} X_{r+s} \nonumber\\
&\qquad
-\frac{5}{2}i\sqrt{\frac{1}{2k}}\big(r-\tfrac{3}{2}\big)\big(r^2-\tfrac{1}{4}\big)\delta_{r+s,0}
-3(GP)_{r+s}
\\
[X_m, K_n] &=
3(n+1)K_{m+n}
+3 i \sqrt{\frac{2}{k}} (m-n) X_{m+n} \nonumber\\&\qquad\qquad
-\frac{5i}{2}\sqrt{\frac{1}{2k}}m(m^2-1)\delta_{m+n,0}
+3(GP)_{m+n} \label{XK}
\end{align}
\begin{align}
[X_n, M_r] &=
\frac{3}{16}\left(3 - 4(n^2 + r^2) + 2n - 4r + 4nr\right)G_{n+r} \nonumber\\&\qquad
+\frac{3 i }{28}\sqrt{\frac{1}{2k}} \left(39-12 (n^2+r^2)+36 n+8 r+32 n r\right)P_{n+r} \nonumber\\&\qquad\qquad
+\frac{1}{4} (15-6 n+14 r)M_{n+r}
+4(GX)_{n+r}
+\frac{4 i}{7}\sqrt{\frac{2}{k}}(PX)_{n+r}
\\
[K_m, K_n] &=
3(m-n) \Bigl(X_{m+n}-L_{m+n}+i \sqrt{\tfrac{2}{k}}K_{m+n}\Bigr)
-\tfrac{7}{2}m(m^2-1)\delta_{m+n,0}
\\
[K_n, M_r] &=
-\frac{1}{4} (n+1) (8 n-22 r-33)P_{n+r}
+3 i \sqrt{\frac{2}{k}} \big(n-r-\tfrac{1}{2}\bigr)M_{n+r}
\nonumber\\&\qquad\qquad
+3(GK)_{n+r}
-6(TP)_{n+r}
\\
\{M_r, M_s\} &=
-\frac{1}{8} \left(20 (r^2+s^2)+46 (r+s)+4 r s+39\right)L_{r+s}
\nonumber\\&\qquad
+\frac{1}{10} \left(22 (r^2+s^2)-40(r+s)-56 r s-93\right)X_{r+s}
\nonumber\\&\qquad
+3 i \big(r+\tfrac{3}{2}\big) \big(s+\tfrac{3}{2}\big) \sqrt{\frac{2}{k}}K_{r+s}
-\frac{35}{24}\big(r^2-\tfrac{9}{4}\big)\big(r^2-\tfrac{1}{4}\big)\delta_{r+s,0}
\nonumber\\&\qquad
-6(GM)_{r+s}
-(KK)_{r+s}
+12(TX)_{r+s}
+\frac{2}{5}(XX)_{r+s}
\end{align}

\subsection{Ramond zero-mode algebra} \label{app:RzeroModeAlgebra}

It proves a good idea to express the Ramond zero-mode algebra in terms of $L_0$, $G_0$, ~~ $\tilde{G}_0=iP_0/\sqrt{15}$, ~ $\tilde{L}_0 = -X_0/5$, as well as the combinations
\begin{equation}
\tilde{K}_0=\frac{iK_0}{\sqrt{15}}+\frac{1}{4}\sqrt{\frac{1}{30k}} \ ,
\end{equation}
and
\begin{equation}
U_0 = -\frac{1}{5}\left(M_0 - \frac{G_0}{4} + i\sqrt{\frac{1}{2k}}P_0\right) \ .
\end{equation}
We use $\approx$ where simplifications of normal ordered products occur on highest weight states. We give only non-vanishing commutators.

\begin{align}
\{G_0, G_0\} &=
2L_{0}-\frac{c}{12}
\\
\{G_0, \tilde{G}_0\} &= \tilde{K}_0
\\
[G_0, \tilde{L}_0] &=U_0
\\
\{\tilde{G}_0, \tilde{G}_0\} &=
2\tilde{L}_0-\frac{7/10}{12}
\\
[\tilde{G}_0, \tilde{K}_0] &= -U_0
\end{align}
\begin{align}
\{\tilde{G}_0, U_0\} &\approx
\frac{3}{10}\left(\tilde{K}_0-2\tilde{G}_0G_0\right)
\\
[\tilde{L}_0, \tilde{K}_0] &\approx -\frac{3}{10}\left(\tilde{K}_0-2\tilde{G}_0G_0\right)
\\
[\tilde{L}_0, U_0] &\approx
\frac{3}{100}G_0-\frac{2}{5}U_0-\frac{24}{7}\sqrt{\frac{1}{30k}}\left(\tilde{L}_0-\frac{3}{80}\right)\tilde{G}_0-\frac{4}{5}\tilde{L}_0G_0
\\
[\tilde{K}_0, U_0] &\approx \frac{6}{5}\tilde{G}_0\left(L_0-\frac{c}{24}\right)-\frac{3}{5}\tilde{K}_0G_0
\\
\{U_0, U_0\} &\approx 
-\frac{12}{5}\left(\tilde{L}_0-\frac{7/10}{24}\right)\left(L_0-\frac{c}{24}\right)
+\frac{2}{5}\left(\tilde{L}_0-\frac{7}{16}\right)\left(\tilde{L}_0-\frac{3}{80}\right) \nonumber\\
&\qquad\qquad
-\frac{6}{5}U_0G_0
+\frac{3}{5}\tilde{K}_0^2 \label{eq:U0U0}
\end{align}

\section{Two-dimensional Ramond highest weight representations} \label{app:Rrep2d}

Assuming a representation as follows for the $\mathcal{N}=1$ generators, we solve for the other $2\times 2$ matrices using the Ramond zero-mode algebra in appendix~\ref{app:RzeroModeAlgebra}. We find the following three solutions.

\begin{flalign}
~\text{\underline{Type III$_{\rm R}$ :}} \quad
L_0=&h\qquad
\tilde{L}_0=\frac{3}{80}\qquad
G_0=
\begin{pmatrix}
0 & 1 \\ h-\frac{c}{24} & 0
\end{pmatrix}\qquad
U_0 = 0\\
&\tilde{G}_0 = \pm\frac{1}{2\sqrt{30}}\qquad\qquad
\tilde{K}_0 =
\pm \frac{1}{\sqrt{30}}
\begin{pmatrix}
0 & 1 \\ h-\frac{c}{24} & 0
\end{pmatrix} \nonumber &&
\end{flalign}

\vspace{10pt}

\begin{flalign}
\text{\underline{Type IV$_{\rm R}$ :}} \quad
&L_0=h\qquad
\tilde{L}_0=\frac{3}{80}\qquad
G_0=
\begin{pmatrix}
0 & 1 \\ h -\frac{c}{24} & 0
\end{pmatrix}\qquad
U_0=0\\
\tilde{G}_0=
\pm &\frac{1}{\sqrt{120}}
\begin{pmatrix}
0 & \left(h-\frac{c}{24}\right)^{-1/2} \\
\left(h-\frac{c}{24}\right)^{1/2} & 0
\end{pmatrix}\qquad
\tilde{K}_0 = \pm\frac{1}{\sqrt{30}}\sqrt{h-\frac{c}{24}} \nonumber &&
\end{flalign}

\vspace{10pt}

\begin{flalign}
\text{\underline{Type V$_{\rm R}$ :}} \quad
&L_0=h\qquad
\tilde{L}_0=\frac{7}{16}\qquad
G_0=
\begin{pmatrix}
0 & 1 \\ h -\frac{c}{24} & 0
\end{pmatrix}\qquad
U_0=0\\
\tilde{G}_0=
\pm &\frac{7}{\sqrt{120}}
\begin{pmatrix}
0 & \left(h-\frac{c}{24}\right)^{-1/2} \\
\left(h-\frac{c}{24}\right)^{1/2} & 0
\end{pmatrix}\qquad
\tilde{K}_0 = \pm\frac{7}{\sqrt{30}}\sqrt{h-\frac{c}{24}} \nonumber &&
\end{flalign}

\section{The analysis for $\text{AdS}_3\times(\text{S}^3\times\mathbb{T}^4)/D_n$}\label{app:Dn}

In this Appendix we show that our construction of Section~\ref{sec:T4} can also be generalised to backgrounds $\text{AdS}_3\times(\text{S}^3\times\mathbb{T}^4)/D_n$ with $\mathcal{N}=(2,2)$ supersymmetry \cite{Datta:2017ert}, see also \cite{ Gaberdiel:2019wjw}. These backgrounds involve orbifolding the world-sheet CFT associated to 
\be
\widehat{\mathfrak{su}}(2)_{k-2} \oplus \widehat{\mathfrak{u}}(1)^{4} \oplus \text{Fer}^7 \ , 
\ee
where the $\mathbb{T}^4$ factor contributes $\widehat{\mathfrak{u}}(1)^{4} \oplus \text{Fer}^4$, 
by the dihedral group $D_n$ with $n=1, 2, 3, 4$ or $6$. The reflection generator of $D_n$ rotates the S$^3$ by 180 degrees in order to break target space isometries enough to achieve the desired amount of supersymmetry in the dual theory. Rotations in $D_n$ act trivially on S$^3$.

In order to allow for a uniform description, we denote the $\mathbb{T}^4$ generators by 
\begin{equation} \label{eq:T4fields}
\begin{aligned}
&\mathcal{K}^i\text{ (bosonic)} & \qquad \mathcal{K}^i(z)\mathcal{K}^j(w) \sim \frac{k }{2 (z-w)^2} \, \delta^{ij}
\\
&\chi^i\text{ (fermionic)} & \qquad \chi^i(z)\chi^j(w) \sim \frac{k }{2 (z-w)^2} \, \delta^{ij}
\end{aligned}
\qquad\quad \text{for }i, j\in\{4,5,6,7\}\ .
\end{equation}
The $D_n$ action on $\mathbb{T}^4$ is naturally inherited from
\begin{equation}
D_n \subset \text{O}(2)_{\text{diag}} \subset \text{O}(2)\times \text{O}(2) \,.
\end{equation}
Concretely let us take $\text{O}(2)\times \text{O}(2)$ to act on $\mathbb{T}^4$ fermions according to
\begin{align}
\begin{pmatrix}
\chi^4 \\ \chi^5
\end{pmatrix}
&\longmapsto
\begin{pmatrix}
\cos\theta_1 & -\sin\theta_1 \\
\epsilon_1\sin\theta_1 & \epsilon_1\cos\theta_1
\end{pmatrix}
\begin{pmatrix}
\chi^4 \\ \chi^5
\end{pmatrix}\ ,
\qquad~~\,
\theta_1 \in [0,2\pi)\ ,~~\epsilon_1=\pm 1\ , \label{eq:O2*O2A}
\\
\begin{pmatrix}
\chi^6 \\ \chi^7
\end{pmatrix}
&\longmapsto
\begin{pmatrix}
\cos\theta_2 & \sin\theta_2 \\
-\epsilon_2\sin\theta_2 & \epsilon_2\cos\theta_2
\end{pmatrix}
\begin{pmatrix}
\chi^6 \\ \chi^7
\end{pmatrix}\ ,
\qquad
\theta_2 \in [0,2\pi)\ ,~~\epsilon_2=\pm 1\ . \label{eq:O2*O2B}
\end{align}
The action on the bosons is identical to the action on the fermions.
The $\epsilon_i$ signs allow us to reach the two disconnected components of
\begin{equation}
\text{O}(2)=\text{SO}(2)\cup S \cdot \text{SO}(2)
\end{equation}
via reflection $S:\theta_i = 0,~\epsilon_i=-1$. $\text{O}(2)_{\text{diag}}$ above is defined by $\theta_1=\theta_2$, $\epsilon_1=\epsilon_2$. 

In terms of the analysis of Section~\ref{sec:T4} it therefore remains to check that the ${\cal N}=1$ field $G$ in (\ref{N1G}), as well as the field $P$ in eq.~(\ref{eq:PforK3}) are invariant under the $D_n$ action. Here we have used that the ${\cal N}=1$ stress-tensor (\ref{N1L}) is generated by the OPE of $G$ with itself $G$, and that the rest of the algebra is obtained from the OPEs of these fields. 
With the above conventions the relevant fields take the form 
\begin{align}
G &= -\frac{4i}{k^2}\chi^1\chi^2\chi^3 + \sum_{i=1}^7 \frac{2}{k} {\mathcal{K}^i\chi^i}\ , \label{GT4}\\
P &=\sqrt{\frac{8}{k^3}}\Bigl({\chi^1\chi^2\chi^3}+\chi^1({\chi^4\chi^5}+{\chi^6\chi^7})+\chi^2({\chi^4\chi^6}-{\chi^5\chi^7})-\chi^3({\chi^4\chi^7}+{\chi^5\chi^6})\Bigr)\ . 
\nonumber
\end{align}
Let us first analyse the behaviour under $S$ ($\theta_i=0$, $\epsilon_i=-1$), for which $\chi^5$ and $\chi^7$, as well as the corresponding bosons, change sign. It is easy to see that $G$ and $P$ remain invariant provided we rotate the S$^3$ fermions by 180 degrees,
\begin{equation}
\chi^1 \longmapsto -\chi^1\,,\qquad
\chi^2 \longmapsto \chi^2\,,\qquad
\chi^3 \longmapsto -\chi^3\ , 
\end{equation}
and similarly for the ${\rm S}^3$ bosons. 

A similar statement can be made for the $\text{AdS}_3\times (\text{S}^3\times \mathbb{T}^2)/\mathbb{Z}_2\times \mathbb{T}^2$ background discussed in section 2.1 of \cite{Datta:2017ert}, for which the relevant $\mathbb{Z}_2$ action is described by $\theta_1=\pi$, $\theta_2=0$, $\epsilon_i=1$, for which $\chi^4$ and $\chi^5$ (as well as the corresponding bosons) change sign. This leaves $G$ and $P$ invariant provided that now $\chi^2$ and $\chi^3$ change sign.
\smallskip

Next we turn to the rotations. The action of $\text{O}(2)\times \text{O}(2)$ on fermion bilinears, see eqs.~\eqref{eq:O2*O2A} and \eqref{eq:O2*O2B}, decomposes into two singlets 
\begin{align}
\chi^4\chi^5 &\longmapsto \epsilon_1 \chi^4\chi^5\ ,
\\
\chi^6\chi^7 &\longmapsto \epsilon_2 \chi^6\chi^7\ ,
\end{align}
as well as a 4-dimensional representation
{\small
\begin{equation*}
\begin{pmatrix}
\chi^4\chi^6\\
\chi^4\chi^7\\
\chi^5\chi^6\\
\chi^5\chi^7
\end{pmatrix}
\mapsto
\begin{pmatrix}
\cos\theta_1\cos\theta_2 &
\cos\theta_1\sin\theta_2 &
-\sin\theta_1\cos\theta_2 &
-\sin\theta_1\sin\theta_2
\\
-\epsilon_2\cos\theta_1\sin\theta_2 &
\epsilon_2\cos\theta_1\cos\theta_2 &
\epsilon_2\sin\theta_1\sin\theta_2 &
-\epsilon_2\sin\theta_1\cos\theta_2
\\
\epsilon_1\sin\theta_1\cos\theta_2 &
\epsilon_1\sin\theta_1\sin\theta_2 &
\epsilon_1\cos\theta_1\cos\theta_2 &
\epsilon_1\cos\theta_1\sin\theta_2
\\
-\epsilon_1\epsilon_2\sin\theta_1\sin\theta_2 &
\epsilon_1\epsilon_2\sin\theta_1\cos\theta_2 &
-\epsilon_1\epsilon_2\cos\theta_1\sin\theta_2 &
\epsilon_1\epsilon_2\cos\theta_1\cos\theta_2
\end{pmatrix}
\begin{pmatrix}
\chi^4\chi^6\\
\chi^4\chi^7\\
\chi^5\chi^6\\
\chi^5\chi^7
\end{pmatrix} \,.
\end{equation*}
}

The two singlets are summed together in our ansatz for $P$. Under a rotation ($\theta_1=\theta_2$, $\epsilon_1=\epsilon_2=1$), we learn that the S$^3$ fermion $\chi^1$ should be invariant, consistently with \cite{Datta:2017ert}. Moreover also
\begin{equation}
\chi^4\chi^6-\chi^5\chi^7\ ,\qquad \hbox{and} \qquad 
\chi^4\chi^7+\chi^5\chi^6\ 
\end{equation}
are invariant. These multiply respectively $\chi^2$ and $\chi^3$ in our ansatz for $P$. Hence, $P$, $G$ and thus all generators of $\mathcal{SW}_k(\frac{3}{2}, \frac{3}{2}, 2)$ are invariant under rotations provided that the S$^3$ generators are invariant. This proves the claim.


\normalem

\end{document}